# Continual Model of Medium III: Calculating the Analytical Gradients of the Parameters of Molecular Surface Elements over the Atomic Coordinates

O. Yu. Kupervasser[a,*] and N. E. Wanner[b]

[a]TRANSIST VIDEO Ltd., Skolkovo resident

[b]All-Russia Research Institute of Veterinary Sanitary, Hygiene, and Ecology, Russian Academy of Agricultural Sciences, Moscow

*e-mail: olegkup@yahoo.com

**Abstract**—The problem of finding the analytical gradients (derivatives over the atomic coordinates) of the solvation energy can be divided into the two subproblems: at the first stage, we search for the parameters of surface elements (three coordinates, three components of the normal unit vector, and surface area) and their derivatives and, at the second stage, we differentiate the energy and express it through the derivatives of the matrices describing the problem. The derivatives of matrix elements are expressed through the derivatives of the parameters of solvent excluded surface (SES) or solvent accessible surface (SAS) elements. The objective of our work is to find these very analytical gradients of the parameters of surface elements.

## 1. Introduction

The objective of our work is to find the analytical gradients of the parameters of SES or SAS elements. The analytical gradients are considered to mean the partial derivatives of parameters over the coordinates of atoms that compose a molecule. These gradients need to be found with the purpose of calculating the analytical gradients of the free solvation energy. These gradients may further be used when searching for the global energy minimum, which is necessary, for example, in the computer-aided modeling of medicines.

Before starting the search for the gradients of surface elements, we need to define for what a surface we perform this search. There exist the two types of surfaces over a molecule. First, this is the solvent excluded surface (SES). The volume occupied by a solvent lies *outside* the volume enveloped by this surface. The substrate itself lies completely *inside* this volume. Second, this is the solvent accessible surface (SAS) that is formed by the centers of solvent molecules tangent to a substrate molecule. The first type of surface is used to calculate the polarization component of the solvation energy, and the second type of surface is used to determine the cavitation and van der Waals components [1].

The earlier performed studies on the analytical gradients of the parameters of SES elements deal with the algorithm implemented in the GEPOL software [2], where the cavities in a molecule are filled with fictitious spheres. For a smooth surface obtained by primary and secondary rolling using the algorithm described in [3–10], such a study is performed in details in our work for the first time. Some algorithms of constructing a smooth surface, a method of its triangulation and partitioning into surface elements (Fig. 1), and some methods of finding their parameters and gradients have already been described in the papers [3–10]. In this work, we shall enlarge on finding the analytical gradients. The calculations of the analytical gradients of surface elements obtained by primary and secondary rolling [3–10] were implemented in the DISOLV software [3–4, 8–10].

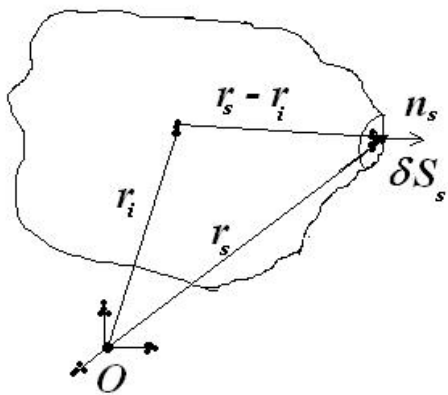

Fig. 1. Surface element relative to the origin of coordinates.

## 2. Analytical Gradients of the Parameters of Surface Elements

Let us find the formulas describing the derivatives of the parameters of SES and SAS elements (coordinates, normals, surface areas) over the atomic coordinates. We shall first consider SES. We have the two types of segments on this surface: fragments of spheres and tori. Let us consider how the parameter of a surface element that completely lies on one such element are changed upon the shift of atoms.

### 2.1. SES Element Lies on a Sphere (Fig. 2)

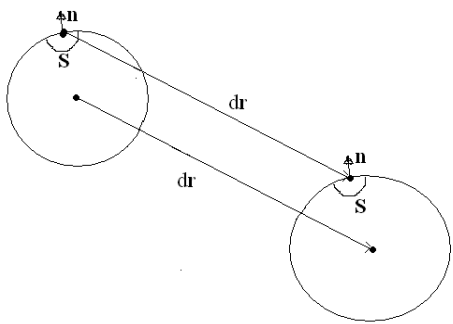

**Fig. 2.** Change in the parameters of a surface element upon the shift of a sphere. It can be seen that only its coordinates are changed, and their changes

are equal to the shift of the center of a sphere. The surface area and normal are constant.

The changes of the coordinates of surface elements are equal to the shift of the center of a sphere, and the surface area and normal are constant.

$$\frac{\partial r_s}{\partial R_{i,x^k}} = \frac{\partial r_m}{\partial x_i^k},$$
$$\frac{\partial n_s}{\partial R_{i,x^k}} = 0,$$
$$\frac{\partial S_s}{\partial R_{i,x^k}} = 0,$$
(1)

where $n_s$ is the normal of a surface element, $S_s$ is the area of a surface element, $r_s$ is the radius vector of the coordinates of a surface element, $r_m$ is the center of a spherical segment, $R_{i,x^k}$ is the $k^{th}$ component of the radius vector $R_i$ ($k=1, 2, 3$), $i$ is the number of a shifted atom ($0 < i < N$), and $N$ is the number of atoms.

## 2.2. SES Element Lies on a Torus

The gradients of the parameters of surface elements are the linear superposition of gradients for the two following cases.

### 2.2.1. One of the Supporting Spheres ($r_2$) is Shifted Perpendicularly to the Axis of a Torus (Fig. 3)

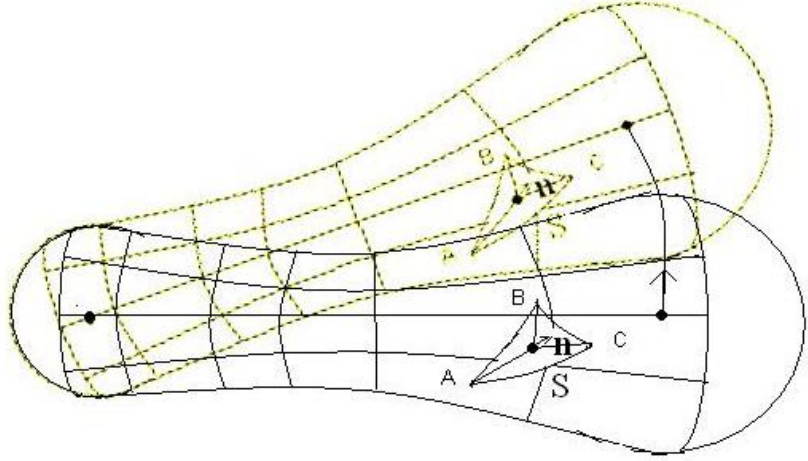

Fig. 3. Rotation of a torus together with a surface element that corresponds to the vertical displacement of the center of one of supporting spheres. The area of a surface element remains unchanged, and its center and normal are rotated through the same angle.

Such a small shift is equivalent to the simple rotation of the entire system of two spheres and a torus around the center of a stationary supporting sphere $r_1$. The center of a surface element rotates through the same angle around the same axis, and its normal also rotates through the same angle. The surface area remains unchanged.

$$\frac{\partial \boldsymbol{\varphi}_s}{\partial R_{i,x^k}} = \frac{\left[(r_2 - r_1) \times \frac{\partial (r_2 - r_1)}{\partial x_i^k}\right]}{(r_2 - r_1)^2},$$

$$\frac{\partial (r_s - r_1)}{\partial R_{i,x^k}} = \left[\frac{\partial_s \boldsymbol{\varphi}}{\partial x_i^k} \times (r_s - r_1)\right], \qquad (2)$$

$$\frac{\partial \boldsymbol{n}_s}{\partial R_{i,x^k}} = \left[\frac{\partial \boldsymbol{\varphi}_s}{\partial x_i^k} \times \boldsymbol{n}_s\right],$$

$$\frac{\partial S_s}{\partial R_{i,x^k}} = 0,$$

where $r_1$ is the coordinate of a stationary atom, $r_2$ is the coordinate of a rotating atom, and $\boldsymbol{\varphi}_s$ is the rotation angle vector.

### 2.2.2. One of the Supporting Spheres ($r_2$) is Shifted

## along the Axis of a Torus (Fig. 4)

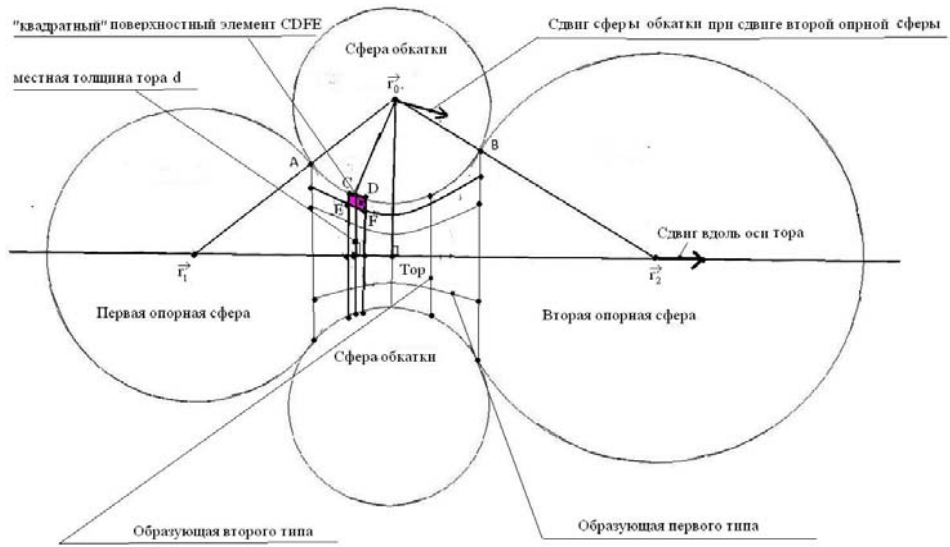

Fig. 4. Rolling torus and two supporting spheres. There is no displacement of elements relative to a rolling sphere upon its forward shift that occurs upon the shift of the second supporting sphere along the axis of a torus. The torus itself becomes thiner (the length of second-type generatrices is reduced), and the length of first-type generatrices is increased due to the elongation of boundary surface elements. The length of first-type generatrix segments remains unchanged for surface elements that completely lie on the torus. The number of surface elements remains constant.

Key:

"квадратный" поверхностный элемент CDFE --> "square" surface element CDEF;

местная толщина тора *d* --> local torus thickness *d*;

Сфера обкатки --> Rolling sphere;

Сдвиг сферы обкатки при сдвиге второй опорной сферы --> Shift of a rolling sphere upon the shift of the second suporting sphere;

Первая опорная сфера --> First supporting sphere;

Вторая опорная сфера --> Second supporting sphere;

Сдвиг вдол оси тора --> Shift along the axis of a torus;

Образующая первого типа --> First-type generatrix;

Образующая второго типа --> Second-type generatrix

At a small shift of the second supporting sphere (see Fig. 4) away from the first supporting sphere along the axis of a torus, this rolling torus is elongated. In this case, there occurs the forward displacemnent (without rotation) of a torus rolling sphere. At such a shift, the normals do not change their direction, and the displacment of all the points of the *first-type* generatrix (the line of contact between a torus and a rolling sphere) will be equal to the displacenent of the center of a rolling sphere. Hence, the gradients of the normal and centers of surface elements are calculated as

$$\frac{\partial r_s}{\partial R_{i,x^k}} = \frac{\partial r_0}{\partial x_i^k},$$
$$\frac{\partial n_s}{\partial R_{i,x^k}} = 0,$$
(3)

where $r_0$ is the displacement of the center of a rolling sphere.

Let us consider a small "square" surface element formed by two segments of first-type torus generatrices (already defined above) and two segments of second-type torus generatrices. Second-type torus generatrices are circles formed by the intersection of a torus and planes that are perpendicular to the axis of this torus. In the case of the forward displacement of a torus rolling sphere, the length of segments that lie on a first-type generatrix remains unchanged. The length of segments belonging to second-type generatrices is reduced due to the "thinning" of a torus in the same proportion as the local thickness of a torus. As a result, the "square" surface element's area equal to the product of the length of a first-type generatrix segment and the length of a second-type generatrix segment is also changed in the same proportion, i.e.,

$$\frac{\partial S_s}{\partial R_{i,x^k}} = \frac{\frac{\partial d}{\partial R_{i,x^k}}}{d} \cdot S_s,$$
(4)

where $d$ is the local thickness of a torus. Any small surface element may be splitted into a set of small "square" elements. Hence, Eq. (45) is also true for this element.

### 2.3. SES Element Lies on the Boundary between Segments of Different Types

During the forward displacement of a torus rolling sphere, a first-type gemeratrix is simulatneously elongated. However, in this case, the length of any first-type gemeratrix segment that completely lay on a torus before displacement remains unchanged. What is a reason for its elongation? Here, it is important to recollect that the number of surface elements is fixed. During the forward displacement of a torus rolling sphere, *boundary* surface elements are elongated. Boundary elements are elements, whose apices lie on segments of different types, i.e., on both a torus and a sphere. The change of their surface areas gives us the sought elongation of first-type generatrices. Using the formula for the surface area of a triangle with three points ($S_s = \frac{1}{2}|(r_1 - r_2) \times (r_3 - r_2)|$) and then taking the gradient, we shall express the gradient of the area of a boundary surface elements through SES point coordinate gradients that are substituted through the above found formulas (42)–(44). The gradients of its normal and coordinates are determined via the same formulas depending on the type of a segment, on which the center of a surface element lies.

### 2.4. SAS Element

SAS elements are obtained by displacing SES triangular elements along the normal. Using the formula for the surface area of a triangle with three points once again and then taking the gradient, we shall express the gradient of the area of SAS element through SES point coordinate and normal gradients that are substituted through the above found formulas (42)–(44).

### 2.5. Result

It can be seen that, knowing the displacement of the centers of spheres, we can find the change in the parameters of surface elements, reasoning from the above stated principles and, consequently, determine their gradients.

## 3. Detailed Formulas for the Analytical Gradinets of Segments of Different Types

Below we shall represent some detailed and comprehensive formulas for the analytical gradients of the parameters of surface elements that lie on primary and secondary spheres and tori of different types and also on atoms. These types were characterized in [3–4].

### 3.1. Atom Surface Elements

The derivatives of the parameters of surface elements (coordinates, normals, surface areas) for the points belonging to the surface of an atom with the center coordinates $R_m$ are completely determined via the gradient of the center of an atom, as mentioned above,

$$\frac{\partial R_m}{\partial R_{i,x^k}} = \delta_{im} e^k$$

$$\frac{\partial r_s}{\partial R_{i,x^k}} = \delta_{im} e^k,$$

$$\frac{\partial n_s}{\partial R_{i,x^k}} = 0, \tag{5}$$

$$\frac{\partial S_s}{\partial R_{i,x^k}} = 0,$$

i.e., the points are displaced together with the center of an atom. Here $e^k$ is the basis unit vector in the direction of the coordinate $x^k$ and a member of the set $\{ e^0, e^1, e^2 \}$, and $\delta_{ij}$ is the delta-function.

### 3.2. Surface Elements That Lie on a Primary Probe Ball Supported by Three Atoms (Fig. 5)

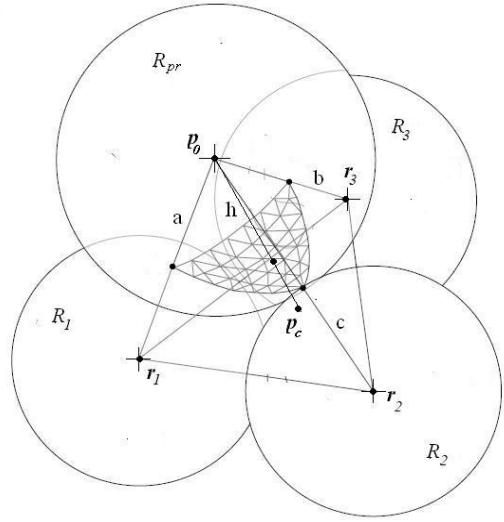

Fig. 5. Rolling in the case of three supporting atoms.

The derivatives of the parameters of surface elements (coordinates, normals, surface areas) that lie on the surface of a primary probe ball supported by three atoms are completely determined by the gradient of the center of this probe ball. For this reason, we first calculate the position and gradients of the center of this probe ball.

Let us determine the position of the center of a probe ball.

We use the following notations in Fig. 5:

$$a = R_1 + R_{pr}; b = R_2 + R_{pr}; c = R_3 + R_{pr}, \qquad (6)$$

where $r$ is the coordinates of the center of a primary rolling sphere $R_{pr}$, $a$, $b$, and $c$ are the edges of a pyramid, $R_1$, $R_2$, and $R_3$ are the radii of supporting spheres, $R_{pr}$ is the radius of a primary rolling sphere, $\mathbf{r}_1 = \mathbf{R}_m, \mathbf{r}_2 = \mathbf{R}_n$, and $\mathbf{r}_3 = \mathbf{R}_l$ are the radii vectors of supporting spheres.

The radius vector that determines the position of a probe ball is the solution of the following set of three vector equations:

$$\begin{cases} (\mathbf{r} - \mathbf{R}_m)^2 = a^2 \\ (\mathbf{r} - \mathbf{R}_n)^2 = b^2 \\ (\mathbf{r} - \mathbf{R}_l)^2 = c^2 \end{cases}, \qquad (7)$$

The radius vector of the center of a probe ball may be resolved into the two vectors: the radius vector of the pyramid base height $p_c$ and the unit vector perpendicular to the base of a pyramid $z$, i.e.,

$$r = p_c \pm zh . \tag{8}$$

These two radii vectors are determined by the two formulas:

$$z = \frac{[(R_l - R_m) \times (R_n - R_m)]}{\|(R_n - R_m) \times (R_l - R_m)\|}, \tag{9}$$

$$p_c = \frac{(R_m + R_n + R_l)}{3} + \frac{1}{6}\left[\begin{array}{l} \frac{3(b^2 - c^2) + (R_l - R_m)^2 - (R_n - R_m)^2}{\|(R_l - R_m) \times (R_n - R_m)\|^2}[[(R_l - R_m) \times (R_n - R_m)] \times R_m] + \\ + \frac{3(c^2 - a^2) + (R_m - R_n)^2 - (R_l - R_n)^2}{\|(R_m - R_n) \times (R_l - R_n)\|^2}[[(R_m - R_n) \times (R_l - R_n)] \times R_n] + \\ + \frac{3(a^2 - b^2) + (R_n - R_l)^2 - (R_m - R_l)^2}{\|(R_n - R_l) \times (R_m - R_l)\|^2}[[(R_n - R_l) \times (R_m - R_l)] \times R_l] \end{array}\right], \tag{10}$$

Hence, the height of a pyramid is found by the formula

$$h^2 = (a^2 + b^2 + c^2)/3 - (R_m^2 + R_n^2 + R_l^2)/3 - p_c^2 + 2p_c \cdot \frac{(R_m + R_n + R_l)}{3}. \tag{11}$$

The derivative of the center of a probe ball is determined as

$$\frac{\partial r}{\partial R_{i,x^k}} = \frac{(r_{x^k} - R_{m,x^k}) \cdot \delta_{im}}{(r - R_m) \cdot [(r - R_n) \times (r - R_l)]}[(r - R_n) \times (r - R_l)] + $$
$$+ \frac{(r_{x^k} - R_{n,x^k}) \cdot \delta_{in}}{(r - R_n) \cdot [(r - R_m) \times (r - R_l)]}[(r - R_m) \times (r - R_l)] + \tag{12}$$
$$+ \frac{(r_{x^k} - R_{l,x^k}) \cdot \delta_{il}}{(r - R_l) \cdot [(r - R_n) \times (r - R_m)]}[(r - R_n) \times (r - R_m)]$$

Hence, the parameters of a surface element are calculated as

$$\frac{\partial r_s}{\partial R_{i,x^k}} = \frac{\partial r}{\partial R_{i,x^k}},$$
$$\frac{\partial n_s}{\partial R_{i,x^k}} = 0, \tag{13}$$
$$\frac{\partial S_s}{\partial R_{i,x^k}} = 0.$$

### 3.3. Surface Elements that Lie on the Sphere of a Secondary Probe Ball Supported by Three Primary Rolling Spheres (Fig. 6)

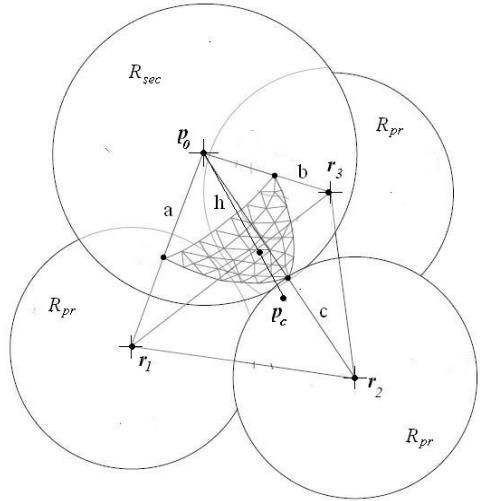

Fig. 6. Rolling in the case of three supporting spheres of primary rolling.

This case is similar to the previous case. It is even simpler, as all the pyramid edges originating from its apex are equal. The radii vectors that determine the position of the supporting spheres $r_1, r_2, r_3$ and the center of a secondary probe ball $r$ form a pyramid. However, the centers of all the three primary supporting spheres may be displaced here upon the displacement of a single atom, so we shall use the principle of superposition over these displacements to find the gradients of parameters.

The radius vector of the center of a probe ball may be resolved into the two vectors: the radius vector of the pyramid base height $p_c$ and the vector perpendicular to the base of a pyramid $z$, i.e.,

$$r = p_c \pm z h. \tag{14}$$

These two radii vectors are determined by the two following formulas:

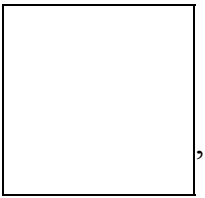

$$\tag{15}$$

$$p_c = \frac{(r_1 + r_2 + r_3)}{3} + \frac{1}{6}\begin{bmatrix} \frac{(r_3 - r_1)^2 - (r_2 - r_1)^2}{\|(r_3 - r_1) \times (r_2 - r_1)\|^2}[[(r_3 - r_1) \times (r_2 - r_1)] \times r_1] + \\ + \frac{(r_1 - r_2)^2 - (r_3 - r_2)^2}{\|(r_1 - r_2) \times (r_3 - r_2)\|^2}[[(r_1 - r_2) \times (r_3 - r_2)] \times r_2] + \\ + \frac{(r_2 - r_3)^2 - (r_1 - r_3)^2}{\|(r_2 - r_3) \times (r_1 - r_3)\|^2}[[(r_2 - r_3) \times (r_1 - r_3)] \times r_3] \end{bmatrix}. \quad (16)$$

The height of a pyramid is determined by the formula

$$h^2 = (R_{pr} + R_{sec})^2 - (r_1^2 + r_2^2 + r_3^2)/3 - p_c^2 + 2 p_c \cdot \frac{(r_1 + r_2 + r_3)}{3}, \quad (17)$$

where $R_{sec}$ is the radius of a secondary rolling sphere.

The displacement of a single atom may lead to the displacement of all the three supporting spheres. The derivative of the center of a probe ball is determined by the displacement superposition, in which each component corresponds to the shift of only one supporting sphere, i.e.,

$$\frac{\partial r}{\partial R_{i,x^k}} = \frac{(r - r_1) \cdot \frac{\partial r_1}{\partial R_{i,x^k}}}{(r - r_1) \cdot [(r - r_2) \times (r - r_3)]}[(r - r_2) \times (r - r_3)] +$$
$$+ \frac{(r - r_2) \cdot \frac{\partial r_2}{\partial R_{i,x^k}}}{(r - r_2) \cdot [(r - r_1) \times (r - r_3)]}[(r - r_1) \times (r - r_3)] + \quad . \quad (18)$$
$$+ \frac{(r - r_3) \cdot \frac{\partial r_3}{\partial R_{i,x^k}}}{(r - r_3) \cdot [(r - r_2) \times (r - r_1)]}[(r - r_2) \times (r - r_1)].$$

The gradients of the centers of primary supporting spheres in this formula are determined via Eq. (13) derived earlier.

Hence, for the parameters of a survface element,

$$\frac{\partial r_s}{\partial R_{i,x^k}} = \frac{\partial r}{\partial R_{i,x^k}},$$
$$\frac{\partial n_s}{\partial R_{i,x^k}} = 0, \quad (19)$$
$$\frac{\partial S_s}{\partial R_{i,x^k}} = 0.$$

### 3.4. Points on the Surface of Secondary Steady-State Position Spheres (Fig. 7)

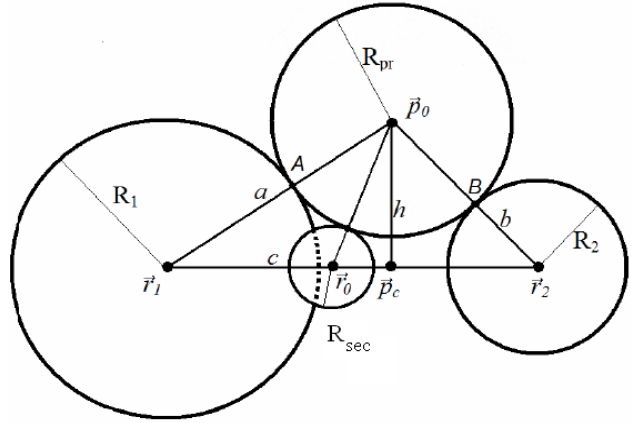

Fig. 7. Secondary steady-state position spheres [10].

The derivatives of the parameters of surface elements (coordinates, normals, surface areas) for the points belonging to the surface of steady-state position spheres are completely determined by the gradient of the centers of these spheres. For this reason, we first calculate the gradients of the centers of these spheres. Let us consider a triangle formed by the centers of two supporting atoms and any of primary rolling spheres:

$$a = R_1 + R_{pr}; \quad b = R_2 + R_{pr}; \quad c = |\mathbf{R}_m - \mathbf{R}_n|, \tag{20}$$

$$a,b - const \quad \frac{\partial a}{\partial R_{i,x^k}} = \frac{\partial b}{\partial R_{i,x^k}} = 0, \tag{21}$$

$$\frac{\partial c}{\partial R_{i,x^k}} = \frac{(\delta_{in} - \delta_{im}) \cdot (R_{n,x^k} - R_{m,x^k})}{c}, \tag{22}$$

where $a$, $b$, and $c$ are the edges of a triangle, $R_{pr}$ is the radius of a primary rolling sphere, $\mathbf{r}_1 = \mathbf{R}_m$, and $\mathbf{r}_2 = \mathbf{R}_n$ are the radii vectors of supporting atoms.

The height of a triangle and its gradient are found as

$$h = \frac{1}{2c}\sqrt{4a^2c^2 - (a^2 + c^2 - b^2)^2}, \tag{23}$$

$$\frac{\partial h}{\partial R_{i,x^k}} = \left(\frac{b^2 + a^2 - c^2}{2h} - h\right) \cdot \frac{\frac{\partial c}{\partial R_{i,x^k}}}{c}. \tag{24}$$

The distance from the center of a steady-state position sphere to the triangle height base point and its gradient are calculated as

$$d = \sqrt{(R_{pr} + R_{sec})^2 - h^2}, \qquad (25)$$

$$\frac{\partial d}{\partial R_{i,x^k}} = -\frac{h \cdot \dfrac{\partial h}{\partial R_{i,x^k}}}{d}, \qquad (26)$$

where $R_{sec}$ is the radius vector of a secondary rolling sphere.

The basis unit vector along the axis of the torus of rolling from the first atom to the second atom and its gradient are determined as

$$z = \frac{(R_n - R_m)}{c}, \qquad (27)$$

$$\frac{\partial z}{\partial R_{i,x^k}} = \frac{(\delta_{in} - \delta_{im})e^k}{c} - \frac{(R_n - R_m)}{c^2}\frac{\partial c}{\partial R_{i,x^k}}, \quad \left(\frac{\partial z}{\partial R_{i,x^k}} \cdot z\right) = 0. \qquad (28)$$

The radius vector of the triangle height base point $p_c$ and its gradient are found as

$$p_c = \tfrac{1}{2}(R_m + R_n) + \frac{(R_n - R_m)(a^2 - b^2)}{2c^2}, \qquad (29)$$

$$\frac{\partial p_c}{\partial R_{i,x^k}} = \tfrac{1}{2}(\delta_{in} + \delta_{im})e^k + \frac{(\delta_{in}-\delta_{im})(a^2-b^2)}{2c^2}e^k - (R_n - R_m)(a^2-b^2)\frac{((\delta_{in}-\delta_{im}) \cdot (R_{n,x^k} - R_{m,x^k}))}{c^4}. \qquad (30)$$

where $r_0$ is the radii vectors of the centers of two steady-state position spheres and their gradients,

$r_0 = p_c \pm dz$, and

$$\frac{\partial r_0}{\partial R_{i,x^k}} = \frac{\partial p_c}{\partial R_{i,x^k}} \pm \left(\frac{\partial d}{\partial R_{i,x^k}} \cdot z + d \cdot \frac{\partial z}{\partial R_{i,x^k}}\right). \qquad (31)$$

Hence, for the parameters of a surface element,

$$\frac{\partial r_s}{\partial R_{i,x^k}} = \frac{\partial r_0}{\partial R_{i,x^k}},$$

$$\frac{\partial n_s}{\partial R_{i,x^k}} = 0, \qquad (32)$$

$$\frac{\partial S_s}{\partial R_{i,x^k}} = 0.$$

## 3.5. Primary Rolling of Two Atoms. Primary Rolling Tori (Fig. 8)

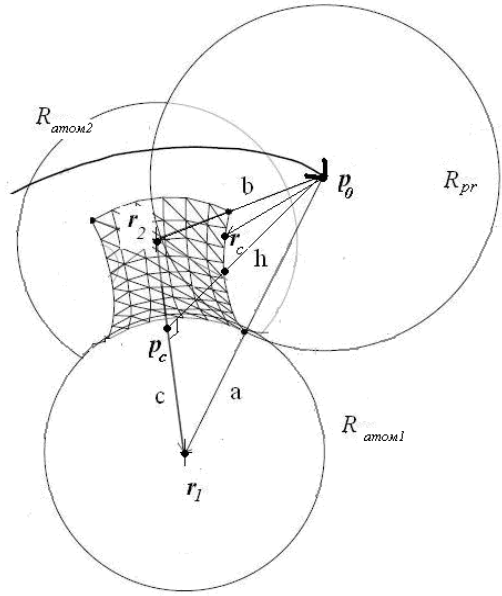

Fig. 8. Primary rolling tori in the case of two supporting atoms.

All the necessary formulas have already been derived in the previous part.

Let $x, y, z$ be the local basis of a torus (the coordinate center lies at the height base point). The $z$ axis is determined as described in the previous part.

$$(x \cdot z)=0, \ (y \cdot z)=0, \ (x \cdot y)=0, \ (x \cdot x)=1, \ (y \cdot y)=1, \ (z \cdot z)=1. \tag{33}$$

The normal of a surface element is expanded in terms of the local basis as

$$n_s = \alpha_s \sqrt{1-\gamma_s^2}\, x + \beta_s \sqrt{1-\gamma_s^2}\, y + \gamma_s z, \tag{34}$$

$$\alpha_s^2 + \beta_s^2 = 1, \tag{35}$$

$$\frac{\partial \gamma_s}{\partial R_{i,x^k}} = 0, \ \frac{\partial \alpha_s}{\partial R_{i,x^k}} = 0, \ \frac{\partial \beta_s}{\partial R_{i,x^k}} = 0. \tag{36}$$

Specifying $\alpha_s$, $\beta_s$, and $\gamma_s$, we determine the position of the center of a surface element on a torus. Here $e_s$ is the unit vector from the height base point to the center of a rolling sphere determined by current $\alpha_s$ and $\beta_s$ as

$$e_s = \alpha_s x + \beta_s y. \tag{37}$$

Its gradient is found as

$$\frac{\partial \boldsymbol{e}_s}{\partial R_{i,x^k}} = -\left(\boldsymbol{e}_s \cdot \frac{\partial \boldsymbol{z}}{\partial R_{i,x^k}}\right)\boldsymbol{z}. \qquad (38)$$

Hence, the gradient of the normal of a surface element for primary rolling tori is determined as

$$\frac{\partial \boldsymbol{n}_s}{\partial R_{i,x^k}} = \sqrt{1-\gamma_s^2}\,\frac{\partial \boldsymbol{e}_s}{\partial R_{i,x^k}} + \gamma_s \cdot \frac{\partial \boldsymbol{z}}{\partial R_{i,x^k}}. \qquad (39)$$

For the gradient of the radius vector of a surface element,

$$\frac{\partial \boldsymbol{r}_s}{\partial R_{i,x^k}} = \frac{\partial \boldsymbol{p}_c}{\partial R_{i,x^k}} + \frac{\partial h}{\partial R_{i,x^k}}\cdot \boldsymbol{e}_s + \left(h - R_{rp}\sqrt{1-\gamma_s^2}\right)\cdot \frac{\partial \boldsymbol{e}_s}{\partial R_{i,x^k}} - R_{rp}\gamma_s \cdot \frac{\partial \boldsymbol{z}}{\partial R_{i,x^k}}. \qquad (40)$$

For the gradient of the areas of surface elements of primary rolling tori,

$$\frac{\partial S_s}{\partial R_{i,x^k}} = \frac{\dfrac{\partial h}{\partial R_{i,x^k}}}{(h - R_{rp}\sqrt{1-\gamma_s^2})}\cdot S_s. \qquad (41)$$

### 3.6. Secondary Rolling of Two Primary Rolling Spheres. Secondary Rolling Tori (Fig. 9)

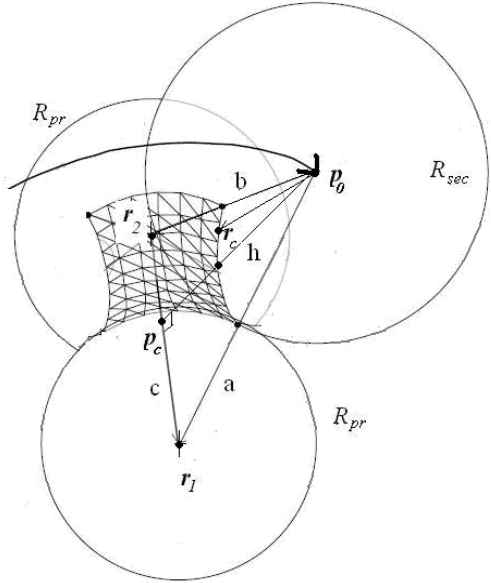

Fig. 9. Secondary rolling tori in the case of two primary supporting spheres.

This case is similar to the previous case. It is even simpler, as supporting spheres have equal radii, and the triangle formed by the centers of two supporting spheres and a primary rolling sphere is equilateral. However, there also exists a distinction: all the normals are oriented in the direction *opposite* to the center of a current rolling sphere. The edges of the triangle formed by the centers of two supporting spheres and a secondary rolling sphere and their gradients are determined as

$$a = b = R_{pr} + R_{sec}; \quad c = |r_2 - r_1|, \tag{42}$$

$$\frac{\partial c}{\partial R_{i,x^k}} = \frac{(\frac{\partial r_2}{\partial R_{i,x^k}} - \frac{\partial r_1}{\partial R_{i,x^k}}) \cdot (r_2 - r_1)}{c}. \tag{43}$$

The height and its gradient are found as

$$h = \frac{1}{2}\sqrt{4a^2 - c^2}, \tag{44}$$

$$\frac{\partial h}{\partial R_{i,x^k}} = -\frac{c}{4h} \cdot \frac{\partial c}{\partial R_{i,x^k}}. \tag{45}$$

The radius vector of the triangle height base point $p_c$ and its derivative are found as

$$p_c = \tfrac{1}{2}(r_1 + r_2),$$
$$\frac{\partial p_c}{\partial R_{i,x^k}} = \tfrac{1}{2}\left(\frac{\partial r_1}{\partial R_{i,x^k}} + \frac{\partial r_2}{\partial R_{i,x^k}}\right). \tag{46}$$

The local basis $x, y, z$ is determined as

$$z = \frac{(r_2 - r_1)}{c}, \quad (x \cdot z) = 0, \quad (y \cdot z) = 0, \quad (x \cdot y) = 0. \tag{47}$$

The derivatives of the local basis are expressed as

$$\frac{\partial z}{\partial R_{i,x^k}} = \frac{\left(\frac{\partial r_2}{\partial R_{i,x^k}} - \frac{\partial r_1}{\partial R_{i,x^k}}\right)}{c} - \frac{(r_2 - r_1)}{c^2}\frac{\partial c}{\partial R_{i,x^k}}. \tag{48}$$

The projection of the normal onto the local basis for secondary rolling tori (the normal is oriented inwards a torus) is determined as

$$n_s = -\alpha_s \cdot \sqrt{1-\gamma_s^2} \cdot x - \beta_s \cdot \sqrt{1-\gamma_s^2} \cdot y - \gamma_s \cdot z. \tag{49}$$

The derivatives of the radius vector of the center of a current rolling sphere, the unit vector from the height base point to the center of a current rolling sphere, and the vectors of the local basis $x, y, z$ are found as

$$e_s = \alpha_s x + \beta_s y, \tag{50}$$

$$\frac{\partial e_s}{\partial R_{i,x^k}} = -\left(e_s \cdot \frac{\partial z}{\partial R_{i,x^k}}\right) z. \tag{51}$$

Hence, the gradient of the normal of a surface element for secondary rolling tori is determined as

$$\frac{\partial n_s}{\partial R_{i,x^k}} = -\sqrt{1-\gamma_s^2}\,\frac{\partial e_s}{\partial R_{i,x^k}} - \gamma_s \cdot \frac{\partial z}{\partial R_{i,x^k}}. \tag{52}$$

For the gradient of the radius vector of a surface element,

$$\frac{\partial r_s}{\partial R_{i,x^k}} = \frac{\partial p_c}{\partial R_{i,x^k}} + \frac{\partial h}{\partial R_{i,x^k}} \cdot e_s + \left(h - R_{sec}\sqrt{1-\gamma_s^2}\right) \cdot \frac{\partial e_s}{\partial R_{i,x^k}} - R_{sec}\gamma_s \cdot \frac{\partial z}{\partial R_{i,x^k}}. \tag{53}$$

For the gradient of the areas of surface elements of secondary rolling tori,

$$\frac{\partial S_s}{\partial R_{i,x^k}} = \frac{\dfrac{\partial h}{\partial R_{i,x^k}}}{(h - R_{sec}\sqrt{1-\gamma_s^2})} \cdot S_s. \tag{54}$$

### 3.7. Consideration of a Boundary Polygonal Surface Element Composed by Triangles, Whose Apices Do Not Belong to the Same Surface Element

The normals and coordinates of the radius vector of a polygonal surface element are determined by the type of a surface fragment, within which its center lies. Using the formulas for this fragment, we calculate both the gradient of the radius vector and normals of this surface element.

It is more difficult to calculate its surface area. It is equal to one third of the total surface area of triangles composing this surface element. Correspondingly, its gradient is calculated as

$$\frac{\partial S^M}{\partial R_{i,x^k}} = \frac{\sum_j \dfrac{\partial S_j^{tr}}{\partial R_{i,x^k}}}{3}. \tag{55}$$

Let a triangle be formed by the apices $r_1, r_2, r_3$.

Then,

$$a_j = r_2 - r_1$$
$$b_j = r_3 - r_1.$$
(56)

The surface area of this triangle is determined as

$$s = [a_j \times b_j],$$
(57)

$$S_j^{tr} = \frac{1}{2}|s|.$$
(58)

The gradient of its surface area is found as

$$\frac{\partial S_j^{tr}}{\partial R_{i,x^k}} = \frac{1}{4} \frac{(\frac{\partial a_j}{\partial R_{i,x^k}} \cdot a_j)b_j^2 + (\frac{\partial b_j}{\partial R_{i,x^k}} \cdot b_j)a_j^2 - (a_j \cdot b_j)((\frac{\partial a_j}{\partial R_{i,x^k}} \cdot b_j) + (a_j \cdot \frac{\partial b_j}{\partial R_{i,x^k}}))}{S_j^{tr}},$$
(59)

$$\frac{\partial a_j}{\partial R_{i,x^k}} = \frac{\partial r_2}{\partial R_{i,x^k}} - \frac{\partial r_1}{\partial R_{i,x^k}},$$
$$\frac{\partial b_j}{\partial R_{i,x^k}} = \frac{\partial r_3}{\partial R_{i,x^k}} - \frac{\partial r_1}{\partial R_{i,x^k}}.$$
(60)

We have supposed above that the gradients of the normals of boundary elements may be calculated from the gradient of the normal to a surface that passes through the center of this element. However, this parameter can be calculated in quite a different way. A polygonal surface element consists of triangular edges. It is possible to calculate the normal as the area-weighed average of the normals of all its edges and, thereupon, to determine its derivative, as performed below. Practical calculations show that these derivatives lead to smaller numerical errors in the gradient.

However, for the energy itself, smaller errors are obtained for the normals taken at the center of a surface element.

This is associated with that polygonal elements consist of triangles belonging to different surface fragments. Their surface areas are considerably changed and apices are appreciably shifted *after* the surface grid is distorted by the shift of the atom, for which the gradient is taken. For this reason, the average normal strongly differs from the normal at the center *after* the surface grid is distorted by the shift of an atom.

Let us describe the above recalculation of the gradient of the normals of boundary surface elements.

The normal of an element is found as the vector average of normals to triangular elements with an apex at the central point of a polygonal element, and the weights are proportional to the surface areas of triangles, i.e.,

$$a_{gi} = r_{gi} - r_g, \tag{61}$$

where $r_g$ is the radius vector of the center of a current surface element, $r_{gi}$ is the radii vectors of adjacent surface elements, and $i=1,...,N_{eg}$, $N_{eg}$ is the number of adjacent surface elements.

Let $n_g$ be the previous normal at the enter of a surface element

$$a_{gi\_next} = \begin{cases} a_{g(i+1)} & \text{for } 1 \leq i < N_{eg} \\ a_{g1} & \text{for } i = N_{eg} \end{cases}. \tag{62}$$

Then a new normal is determined as

$$n_{g(new)} = \frac{\sum_{i=1}^{N_{eg}} \left( [a_{gi} \times a_{gi\_next}] sign(n_g [a_{gi} \times a_{gi\_next}]) \right)}{\left| \sum_{i=1}^{N_{eg}} \left( [a_{gi} \times a_{gi\_next}] sign(n_g [a_{gi} \times a_{gi\_next}]) \right) \right|}. \tag{63}$$

The gradient of the normal is calculated through the gradients of the apices of triangles using the formula similar to the equation for the gradient of the area of a boundary surface element, i.e.,

$$L_{gi} = sign(n_g \cdot [a_{gi} \times a_{gi\_next}]), \tag{64}$$

$$s_g = \sum_i L_{gi} [a_{gi} \times a_{gi\_next}], \tag{65}$$

$$\frac{\partial a_{gi}}{\partial R_{m,x^k}} = \frac{\partial r_{gi}}{\partial R_{m,x^k}} - \frac{\partial r_g}{\partial R_{m,x^k}}, \tag{66}$$

$$\frac{\partial a_{gi\_next}}{\partial R_{m,x^k}} = \frac{\partial r_{gi\_next}}{\partial R_{m,x^k}} - \frac{\partial r_g}{\partial R_{m,x^k}}, \tag{67}$$

$$S_{gi} = \left( \left[ \frac{\partial a_{gi}}{\partial R_{m,x^k}} \times a_{gi\_next} \right] + \left[ a_{gi} \times \frac{\partial a_{gi\_next}}{\partial R_{m,x^k}} \right] \right) L_{gi}, \tag{68}$$

$$\frac{\partial n_{g(new)}}{\partial R_{m,x^k}} = \frac{\sum_{i=1}^{N_{eg}} \left( S_{gi} - n_{g(new)} (n_{g(new)} \cdot S_{gi}) \right)}{|\vec{s}_g|}. \tag{69}$$

## 3. SAS Elements and Their Analytical Gradients

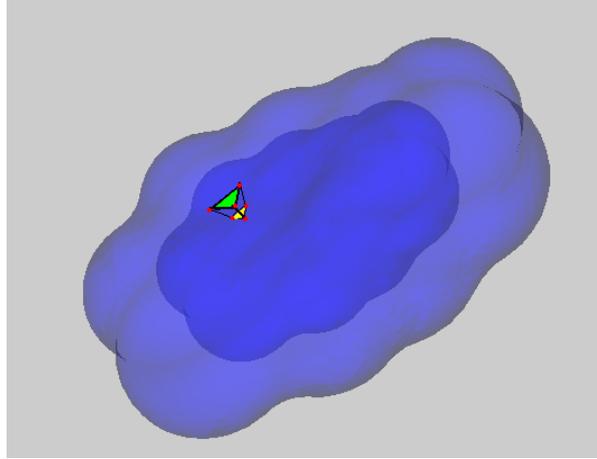

Fig. 10. Transformation of SES into SAS.

Let there be SES after primary rolling. Let there exist a triangle on primary SES ($r_1$, $r_2$, $r_3$). Then the image of this triangle on SAS ($r_{1n}$, $r_{2n}$, $r_{3n}$) (Fig. 10) is determined as

$$r_{1n} = r_1 + n_1 R_{pr}, \tag{70}$$

$$r_{2n} = r_2 + n_2 R_{pr}, \tag{71}$$

$$r_{3n} = r_3 + n_3 R_{pr}, \tag{72}$$

where $R_{pr}$ is the primary rolling radius and $n_1$, $n_2$, and $n_3$ are the normals at corresponding points.

(1) A surface element on a primary sphere supported by three atoms is mapped into a point. Its surface area gradient is zero.

(2) A surface element on a torus supported by two atoms is mapped into a line. Its surface area gradient is also zero.

(3) A surface element on an atom is mapped into a spherical triangle. Its surface area gradient with respect to the shift of atoms is zero.

(4) A boundary surface element between a torus and a primary sphere supported by three atoms has a zero surface area and, consequently, its surface area gradient is zero.

(5) For a boundary triangular surface element supported by a primary torus and an atom, a primary sphere and an atom, or a primary torus, a sphere, and an atom, the surface area is calculated by Eq. (58) for the three points ($r_{1n}, r_{2n}, r_{3n}$). Its surface area gradient is determined as

$$\frac{\partial s_j}{\partial R_{m,x^k}} = \frac{\left((b \cdot b)\left(\frac{\partial a}{\partial R_{m,x^k}} \cdot a\right) + (a \cdot a)\left(\frac{\partial b}{\partial R_{m,x^k}} \cdot b\right) - (a \cdot b)\left(\left(\frac{\partial a}{\partial R_{m,x^k}} \cdot b\right) + \left(\frac{\partial b}{\partial R_{m,x^k}} \cdot a\right)\right)\right)}{S_4}, \quad (73)$$

where the edges of a triangular surface element and their gradients are found as

$$a = r_{2n} - r_{1n}, \quad (74)$$

$$b = r_{3n} - r_{1n}, \quad (75)$$

$$\frac{\partial a}{\partial R_{m,x^k}} = \frac{\partial r_{2n}}{\partial R_{m,x^k}} - \frac{\partial r_{1n}}{\partial R_{m,x^k}}, \quad (76)$$

$$\frac{\partial b}{\partial R_{m,x^k}} = \frac{\partial r_{3n}}{\partial R_{m,x^k}} - \frac{\partial r_{1n}}{\partial R_{m,x^k}}, \quad (77)$$

$$S_4 = 2|a \times b|, \quad (78)$$

and the gradients $r_{1n}, r_{2n}, r_{3n}$ are determined via

$$\frac{\partial r_{1n}}{\partial R_{m,x^k}} = \frac{\partial r_{1n}}{\partial R_{m,x^k}} + \frac{\partial n_1}{\partial R_{m,x^k}} R_{pr}, \quad (79)$$

$$\frac{\partial r_{2n}}{\partial R_{m,x^k}} = \frac{\partial r_{2n}}{\partial R_{m,x^k}} + \frac{\partial n_2}{\partial R_{m,x^k}} R_{pr}, \quad (80)$$

$$\frac{\partial r_{3n}}{\partial R_{m,x^k}} = \frac{\partial r_{3n}}{\partial R_{m,x^k}} + \frac{\partial n_3}{\partial R_{m,x^k}} R_{pr}. \quad (81)$$

The index $j=1,..., N$ runs over all the triangular SAS elements with a non-zero gradient.

## 5. Computer-Aided Verification of Calculations

The computer-aided verification of the above derived formulas was performed using the DISOLV software [4–5, 8–10] as follows:

(1) Two rolling surfaces—the initial surface of a molecule and the surface obtained via the small shift of one of the atoms—were constructed. The shift of grid nodes was calculated through the above found gradients of the coordinates of surface elements, multiplying them by the mentioned shift of an atom. These nodes were then projected onto the surface obtained after the shift of an atom. In this manner, we obtained the two new grids—the shifted grid and its projection onto a new surface. If the gradients are correct, the difference between the positions of nodes of these two grids must be small and much smaller than the difference between them and the initial grid on the initial molecular surface. Moreover, this difference between the two new grids must be reduced linearly (and more quickly than in a linear fashion) with a decrease in the shift of an atom, as it is determined by second derivatives. All these features were really observed in our computer-aided calculations. This confirms the correctness of the formulas for the coordinates of surface elements.

(2) The numerical gradients of the areas of surface elements were found through the difference between the areas of surface elements on the old and new obtained grid. This difference was compared with the surface element area gradient multiplied by the shift. Good agreement of results confirms the correctness of the formulas for the derivatives of the areas of surface elements.

(3) The same procedures as those described in (1) were performed, but, in addition to them, the new projected grid was shifted by the value of the normal to the new surface. The new grid obtained before projection was also shifted by the length of the recalculated normal (the old normal plus its gradient multiplied by the shift of an atom). These two new shifted grids were compared with each other by the same principles as those stated in (1).

## 6. Conclusions

In this work, the analytical gradients of the parameters of surface elements of a molecular surface obtained via primary and secondary rolling by the algorithms described in [3–10] were calculated.

Let us list the problem cases, which may arise when calculating the gradients and solving the problem of optimizing the configuration of molecules with the use of these gradients:

- Degeneration, when a triple point is supported by several atoms instead of three atoms (e.g., aromatic rings);

- The sharp reconstruction of a molecular surface due to that the secondary rolling radius or the critical distance are changed by the program itself in the case of their automatic adjustment;
- The sharp reconstruction of a molecular surface due to the appearance or disappearance of a rolling torus (primary or secondary) upon the shift of triple points or the change of a narrow torus neck at distances close to the critical distance; and
- The sharp reconstruction of a molecular surface due to the appearance or disappearance of primary or secondary triple points.

As a result, there may be gradient jumps and optimizer's "oscillations" near the above described points.

The computer-aided calculation gives good agreement with the expected results and confirms the correctness of the formulas for the gradients of the normals of surface elements.

**Acknowledgments**

The given study is a more detailed presentation and further development of the work [10]. We are profoundly grateful to all the authors for their studies taken as the basis for the given paper.

# КОНТИНУАЛЬНАЯ МОДЕЛЬ СРЕДЫ III: ВЫЧИСЛЕНИЕ АНАЛИТИЧЕСКИХ ГРАДИЕНТОВ ПАРАМЕТРОВ ПОВЕРХНОСТНЫХ ЭЛЕМЕНТОВ НА МОЛЕКУЛЯРНЫХ ПОВЕРХНОСТЯХ ПО КООРДИНАТАМ АТОМОВ.

Купервассер* О.Ю., Ваннер** Н.Э.

* ООО «Транзист Видео», участник Сколково

***Государственное научное учреждение Всероссийский научно-исследовательский институт ветеринарной санитарии, гигиены и экологии Россельхозакадемии, Москва*

*E-mail: olegkup@yahoo.com

Задача нахождения аналитических градиентов (производных по координатам атомов) энергии сольватации распадается на две подзадачи: на первом этапе мы ищем параметры поверхностных элементов (три координаты, три компонента единичного вектора нормали и площадь) и их производные, на втором этапе мы дифференцируем энергию и выражаем ее через производные от матриц, описывающих задачу. При этом производные матричных элементов выражаются через производные параметров поверхностных элементов SES (поверхность исключённого из растворителя объёма) или SAS (поверхность доступная растворителю). Целью данной работы является нахождение именно этих аналитических градиентов от параметров поверхностных элементов.

Ключевые слова: поверхностные элементы, аналитические градиенты, численные градиенты, молекулярная поверхность, первичная обкатка, вторичная обкатка

## 1. Введение.

Целью данной работы является нахождение аналитических градиентов параметров поверхностных элементов на поверхности исключенного объема и поверхности доступной растворителю. Под аналитическими градиентами подразумеваются частные производные от параметров по координатам атомов, составляющих молекулу. Нахождение этих градиентов необходимо для нахождения аналитических градиентов свободной энергии сольватации. Эти градиенты могут использоваться в дальнейшем для поиска глобального минимума энергии, что необходимо, например, в компьютерном моделировании лекарств.

Прежде чем искать градиенты поверхностных элементов, необходимо определить для какой поверхности это делается. Существуют два вида поверхности вокруг молекулы. Во-первых, это SES (Solvent Excluded Surface) - поверхность исключённого из растворителя объёма. Объем, занимаемый растворителем, лежит *вне* объема, ограниченного этой поверхностью. Сам субстрат полностью лежит *внутри* этого объема. Во-вторых, это SAS (Solvent Accessible Surface) - поверхность доступная растворителю образуется центрами молекул растворителя, касающихся молекулы субстрата. Первый тип поверхности используется для расчета поляризационной составляющей энергии сольватации, а второй – для расчета кавитационной о ванн-дер-ваальсовской составляющих [1].

Проведенные ранее работы по аналитическим градиентам параметров SES относились к алгоритму, реализованному в программе GEPOL [2], где пустоты в молекуле заполняются фиктивными сферами. Для гладкой поверхности, полученной первичной и вторичной обкаткой алгоритмом, описанным в [3-10], такая работа проделывается в подробностях впервые в настоящей статье. В работах [3-10] уже описаны алгоритмы построения гладкой поверхности, метод ее триангуляции и разбиения на поверхностные элементы (Рис.1) и методы нахождения их параметров и градиентов. Здесь же остановимся подробно на нахождении аналитических градиентов. Расчеты аналитических градиентов поверхностных элементов, полученной методами первичной и вторичной обкатки [3-10] реализованы в программе DISOLV [3-4,8-10]

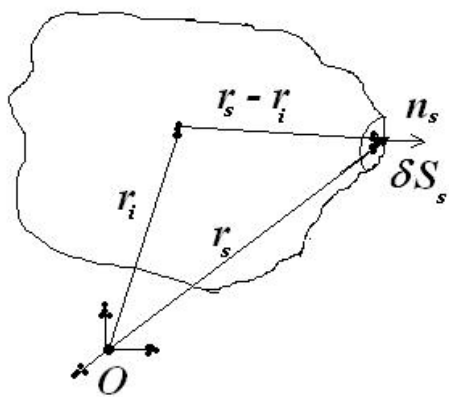

Рис. 1 Элемент поверхности относительно начала координат.

## 2. Аналитические градиенты параметров поверхностных элементов.

Найдем формулы, описывающие производные от параметров поверхностных элементов (координат, нормалей, площадей) по координатам атомов для SES и SAS. Рассмотрим вначале поверхность SES. У нас имеются два типа сегментов этой поверхности – фрагменты сфер и фрагменты торов. Рассмотрим, как меняются при сдвиге атомов параметры поверхностного элемента, полностью лежащего на одном из таких фрагментов.

**2.1 Поверхностный элемент SES лежит на сфере. (Рис.2)**

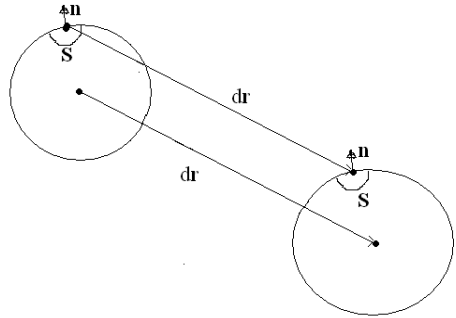

Рис. 2 Изменение параметров поверхностного элемента сферы при сдвиге сферы. Видно, что меняется только его координаты и их изменения равны сдвигу центра сферы. Площадь и нормаль постоянны.

Изменения координат поверхностных элементов равны сдвигу центра сферы. Площадь и нормаль постоянны.

$$\frac{\partial \boldsymbol{r}_s}{\partial R_{i,x^k}} = \frac{\partial \boldsymbol{r}_m}{\partial x_i^k},$$

$$\frac{\partial \boldsymbol{n}_s}{\partial R_{i,x^k}} = 0,$$

$$\frac{\partial S_s}{\partial R_{i,x^k}} = 0,$$

(1)

где $\boldsymbol{n}_s$ - нормаль поверхностного элемента; $S_s$ - площадь поверхностного элемента; $\boldsymbol{r}_s$ - радиус-вектор координат поверхностного элемента; $\boldsymbol{r}_m$ - центр сферического сегмента; $R_{i,x^k}$ - $k$-ая компонента радиус-вектора $R_i$ ($k=1,2,3$); $i$ – номер сдвигаемого атома ($0<i<N$); $N$ число атомов.

### 2.2 Поверхностный элемент SES лежит на торе.

Градиенты параметров поверхностных элементов являются линейной суперпзицией градиентов для следующих двух случаев:

### 2.2.1 Сдвиг одной из опорных сфер ($r_2$) просходит перпендекулярно оси тора.(Рис.3)

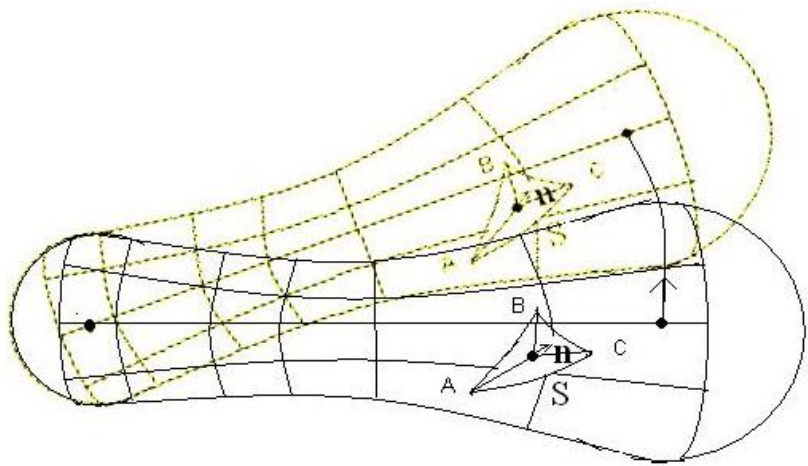

Рис. 3 Поворот тора вместе с поверхностным элементом, соответствующий вертикальному смещению центра одной из опорных сфер. Площадь поверхностного элемента не меняется, его центр и нормаль поворачиваются на тот же угол.

Такой малый сдвиг эквивалентен просто повороту всей системы из 2-ух сфер и тора вокруг центра неподвижной опрорной сферы $r_1$. Центр поверностного элемента поворачивается на тот же угол вокруг той же оси, его нормаль также вращается на тот же угол. Площадь же остается неизменной.

$$\frac{\partial \boldsymbol{\varphi}_s}{\partial R_{i,x^k}} = \frac{\left[(\boldsymbol{r}_2 - \boldsymbol{r}_1) \times \frac{\partial (\boldsymbol{r}_2 - \boldsymbol{r}_1)}{\partial x_i^k}\right]}{(\boldsymbol{r}_2 - \boldsymbol{r}_1)^2},$$

$$\frac{\partial (\boldsymbol{r}_s - \boldsymbol{r}_1)}{\partial R_{i,x^k}} = \left[\frac{\partial_s \boldsymbol{\varphi}}{\partial x_i^k} \times (\boldsymbol{r}_s - \boldsymbol{r}_1)\right],$$

$$\frac{\partial \boldsymbol{n}_s}{\partial R_{i,x^k}} = \left[\frac{\partial \boldsymbol{\varphi}_s}{\partial x_i^k} \times \boldsymbol{n}_s\right],$$

$$\frac{\partial S_s}{\partial R_{i,x^k}} = 0,$$

(2)

$\boldsymbol{r}_1$ - координата неподвижного атома, $\boldsymbol{r}_2$ - координата вращающегося атома, $\boldsymbol{\varphi}_s$ - вектор угла поворота.

### 2.2.2 Сдвиг одной из опорных сфер ($r_2$) просходит вдоль оси тора. (Рис.4)

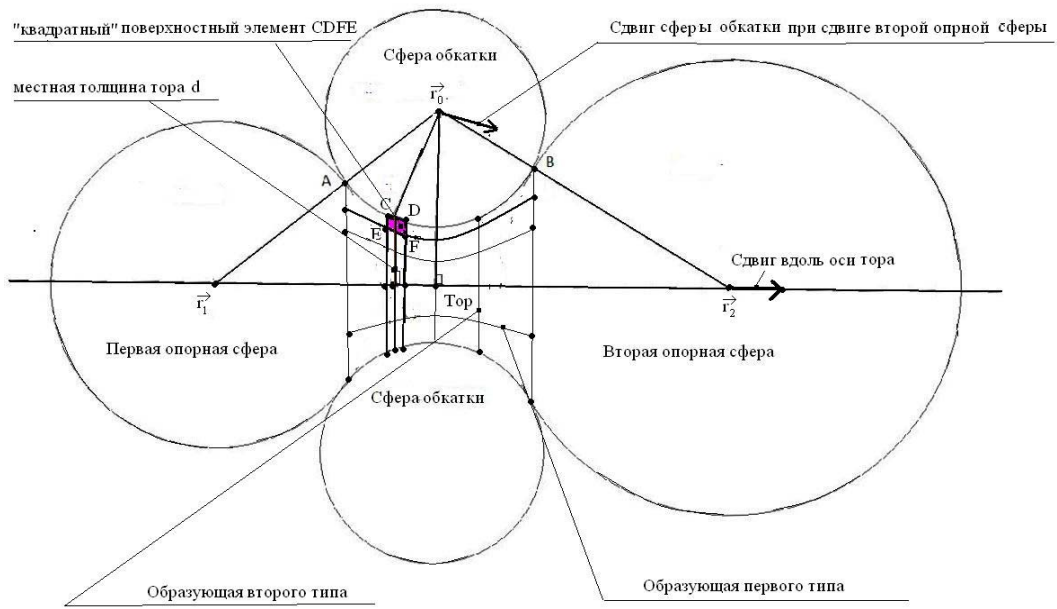

Рис. 4. Тор обкатки и две опорные сферы. Не происходит смещения элементов относительно сферы обкатки при ее поступательном сдвиге, происходящем при сдвиге второй опорной сферы вдоль оси тора. Сам тор «утоньшается» (длина образующих второго типа уменьшается), а длина образующих первого типа возрастает за счет растяжения граничных поверхностных элеметов. При этом длина отрезков образующей первого типа для поверхностных элементов, лежащих полностью на торе, остается неизмененой. Число поверхностных элементов не меняется.

При малом сдвиге второй опорной сферы (см. рис.) в сторону от первой опорной сферы вдоль оси тора этот тор обкатки удлинняется. При этом происходит поступательное (без вращения) смещения сферы обкатки тора. При этом сдвиге нормали не меняют направление, а смещение всех точек образующей *первого* типа (это линия касания тора сферой обкатки) будет такое же, как смещение центра сферы обкатки. Отсюда градиенты нормалей и центров поверхностных элементов вычисляются по формулам:

$$\frac{\partial \boldsymbol{r}_s}{\partial R_{i,x^k}} = \frac{\partial \boldsymbol{r}_0}{\partial x_i^k},$$

$$\frac{\partial \boldsymbol{n}_s}{\partial R_{i,x^k}} = 0,$$

(3)

где $\boldsymbol{r}_0$ – смещение центра сферы обкатки.

Рассмотрим маленький «квадратный» поверхностный элемент, образованный двумя отрезками образующих тора первого типа (уже определены выше) и двумя отрезками образующих тора второго типа. Образующие тора второго типа – это окружности, образованные пересечением плоскостей, перпендикулярных оси тора с самим тором. При поступательном смещении сферы обкатки тора, длина отрезков, лежащих на образующей первого типа, не меняется. Длина отрезков образующих второго типа уменьшается из-за "утоньшения" тора и в той же пропорции, что и местная толщина тора. В итоге и площадь «квадратного» поверхностного элемента (равного произведению длины отрезка образующей первого типа на длину отрезка образующей второго типа) меняется в той же пропорции:

$$\frac{\partial S_s}{\partial R_{i,x^k}} = \frac{\frac{\partial d}{\partial R_{i,x^k}}}{d} \cdot S_s ,$$

(4)

d – местная толщина тора. Любой малый поверхностный элемент можно разбить на набор малых «квадратных» элементов. Следовательно формула (45) верна и для него.

### 2.3 Поверхностный элемент SES лежит на границе между сегментами разного типа.

При поступательном смещения сферы обкатки тора одноременно происходит удлинение образущей первого типа. Однако при этом длины любых отрезков образущей первого типа, лежащие до смещения полностью на торе, не меняются. За счет чего же происходит ее удлиннение? Здесь важно вспомнить, что число поверхностных элементов фиксировано. При поступательном смещения сферы обкатки тора присходит растягивание *граничных* поверхностных элементов. Граничные элементы - это элементы, вершины которых лежит на сегментах разного типа, т.е. как на торе, так и на сфере. Изменение их площади и дает искомое удлинение образующих первого типа. Используя формулу для площади треугольника по трем точкам ($S_s = \frac{1}{2}|(r_1 - r_2) \times (r_3 - r_2)|$), а затем, беря градиент, мы находим выражение для градиента площади граничного элемента поверхности через градиенты координат для точек SES, которые подставляются через уже найденные выше формулы (42-44). Градиенты его нормали и координат определяются теми же формулами в зависимости от типа сегмента, на котором лежит центр поверхностного элемента.

### 2.4 Поверхностный элемент SAS.

Поверхностные элементов на SAS получены сдвигом вдоль нормали треугольных элементов SES. Снова используя формулу для площади треугольника по трем точкам, а затем, беря градиент, мы находим выражение для градиента площади поверхностного элемента на SAS через градиенты координат и нормалей для точек SES, которые подставляются через уже найденные выше формулы (42-44).

**2.5 Итог:**

Мы видим, что зная смещение центров сфер, мы можем найти изменения параметров поверхностных элементов, исходя из изложенных выше принципов, а, следовательно, и найти их градиенты.

## 3. Детальные формулы для аналитических градиентов на различных типах сегментов.

Приведем ниже подробные и детальные формулы для аналитических градиентов параметров поверхностных элементов, лежащих на различных типах первичных и вторичных сфер и торов, а также атомах. Эти типы определены в работах [3-4].

### 3.1 Поверхностные элементы на атомах

Производные от параметров поверхностных элементов (координат, нормалей, площадей) для точек находящихся на поверхности атома с координатами центра $\boldsymbol{R}_m$ полностью определяются через градиент центра атома, как говорилось выше:

$$\frac{\partial \boldsymbol{R}_m}{\partial R_{i,x^k}} = \delta_{im} \boldsymbol{e}^k$$

$$\frac{\partial \boldsymbol{r}_s}{\partial R_{i,x^k}} = \delta_{im}\boldsymbol{e}^k,$$

$$\frac{\partial \boldsymbol{n}_s}{\partial R_{i,x^k}} = 0,$$

$$\frac{\partial S_s}{\partial R_{i,x^k}} = 0,$$

( 5 )

- то есть точки перемещаются вместе с центром атома.

$e^k$ - базисный орт в направлении координаты $x^k$, член множества { $e^0$ , $e^1$ , $e^2$ }),

$\delta_{ij}$ − дельта-функция

### 3.2 Поверхностные элементы, находящихся на сфере первичного шара-зонда, при его опоре на три атома. (Рис.5)

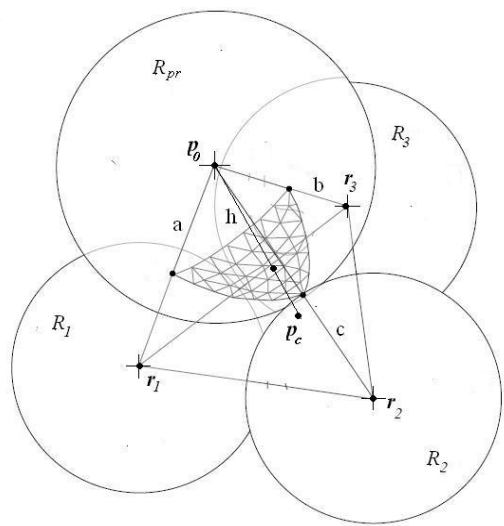

Рис. 5 Обкатка при опоре на три атома.

Производные от параметров поверхностных элементов (координат, нормалей, площадей), находящихся на поверхности сферы первичного шара-зонда и опирающегося на три атома, полностью определяются градиентом центра этого шара-зонда. Поэтому вначале вычислим положение и градиенты центра этого шара-зонда.

Определим положение центра шара-зонда.

Обозначения на Рис. 5:

$$a = R_1 + R_{pr}; b = R_2 + R_{pr}; c = R_3 + R_{pr},$$
( 6 )

$r$ - координаты центра сферы первичной обкатки $R_{pr}$

$a, b, c$ – стороны пирамиды, $R_1, R_2, R_3$ – радиусы опорных сфер, $R_{pr}$ - радиус сферы первичной обкатки; $r_1 = R_m, r_2 = R_n, r_3 = R_l$ - радиус-векторы опорных сфер.

Радиус-вектор, определяющий положение шара-зонда, является решением системы трех векторных уравнений:

$$\begin{cases} (r - R_m)^2 = a^2 \\ (r - R_n)^2 = b^2 \\ (r - R_l)^2 = c^2 \end{cases},$$
( 7 )

Радиус-вектор центра шара-зонда можно разложить по двум векторам - это радиус-вектор основания высоты пирамиды $p_c$ и вектор, перпендикулярный основанию пирамиды $z$:

$$r = p_c \pm z h.$$
( 8 )

Эти два радиус-вектор определяются двумя формулами;

$$z = \frac{[(R_l - R_m) \times (R_n - R_m)]}{\|[(R_n - R_m) \times (R_l - R_m)]\|},$$
( 9 )

$$p_c = \frac{(R_m + R_n + R_l)}{3} + \frac{1}{6}\left[\begin{array}{l}\left(\dfrac{3(b^2 - c^2) + (R_l - R_m)^2 - (R_n - R_m)^2}{\|(R_l - R_m)\times(R_n - R_m)\|^2}[[(R_l - R_m)\times(R_n - R_m)]\times R_m] + \right.\\ +\dfrac{3(c^2 - a^2) + (R_m - R_n)^2 - (R_l - R_n)^2}{\|(R_m - R_n)\times(R_l - R_n)\|^2}[[(R_m - R_n)\times(R_l - R_n)]\times R_n] + \\ \left.+\dfrac{3(a^2 - b^2) + (R_n - R_l)^2 - (R_m - R_l)^2}{\|(R_n - R_l)\times(R_m - R_l)\|^2}[[(R_n - R_l)\times(R_m - R_l)]\times R_l]\right)\end{array}\right], \quad (10)$$

Отсюда высота пирамиды находится по формуле:

$$h^2 = (a^2 + b^2 + c^2)/3 - (R_m^2 + R_n^2 + R_l^2)/3 - p_c^2 + 2p_c \cdot \frac{(R_m + R_n + R_l)}{3}. \qquad (11)$$

Производная центра шара-зонда.

$$\frac{\partial \boldsymbol{r}}{\partial R_{i,x^k}} = \frac{(r_{x^k} - R_{m,x^k})\cdot \delta_{im}}{(\boldsymbol{r} - \boldsymbol{R}_m)\cdot[(\boldsymbol{r} - \boldsymbol{R}_n)\times(\boldsymbol{r} - \boldsymbol{R}_l)]}[(\boldsymbol{r} - \boldsymbol{R}_n)\times(\boldsymbol{r} - \boldsymbol{R}_l)] +$$
$$+ \frac{(r_{x^k} - R_{n,x^k})\cdot \delta_{in}}{(\boldsymbol{r} - \boldsymbol{R}_n)\cdot[(\boldsymbol{r} - \boldsymbol{R}_m)\times(\boldsymbol{r} - \boldsymbol{R}_l)]}[(\boldsymbol{r} - \boldsymbol{R}_m)\times(\boldsymbol{r} - \boldsymbol{R}_l)] + \qquad (12)$$
$$+ \frac{(r_{x^k} - R_{l,x^k})\cdot \delta_{il}}{(\boldsymbol{r} - \boldsymbol{R}_l)\cdot[(\boldsymbol{r} - \boldsymbol{R}_n)\times(\boldsymbol{r} - \boldsymbol{R}_m)]}[(\boldsymbol{r} - \boldsymbol{R}_n)\times(\boldsymbol{r} - \boldsymbol{R}_m)]$$

Отсюда для параметров поверхностного элемента:

$$\frac{\partial \boldsymbol{r}_s}{\partial R_{i,x^k}} = \frac{\partial \boldsymbol{r}}{\partial R_{i,x^k}},$$
$$\frac{\partial \boldsymbol{n}_s}{\partial R_{i,x^k}} = 0,$$
$$\frac{\partial S_s}{\partial R_{i,x^k}} = 0.$$

(13)

### 3.3 Поверхностные элементы, находящихся на сфере вторичного шара-зонда, при его опоре на три сферы первичной обкатки. (Рис.6)

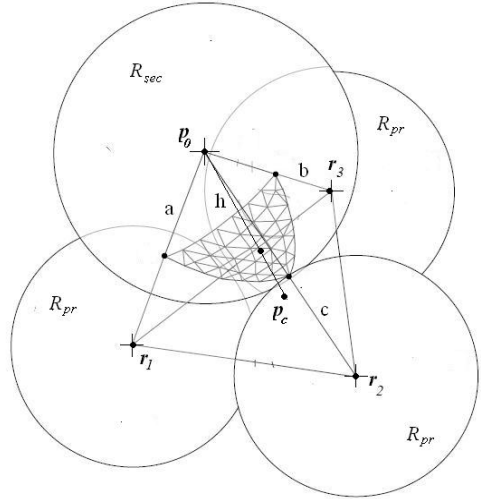

Рис. 6 Обкатка при опоре на три сферы первичной обкатки.

Этот случай похож на предыдущий. Он даже проще, поскольку все ребра пирамиды, выходящие из ее вершины, равны. Радиус-вектора, определяющие положения опорных сфер $r_1$, $r_2$, $r_3$ и центр вторичного шара-зонда $r$ образуют пирамиду. Однако смещаться при смещении одного атома могут центры уже всех трех первичных опорных сфер, поэтому для нахождения градиентов параметров используем принцип суперпозиции по этим смещениям.

Радиус-вектор центра шара-зонда можно разложить по двум векторам - это радиус-вектор основания высоты пирамиды $p_c$ и вектор, перпендикулярный основанию пирамиды $z$:

$$r = p_c \pm zh.$$
( 14 )

Эти два радиус-вектор определяются двумя формулами

$$z = \frac{[(r_3 - r_1) \times (r_2 - r_1)]}{\|[(r_2 - r_1) \times (r_3 - r_1)]\|},$$

(15)

$$p_c = \frac{(r_1 + r_2 + r_3)}{3} + \frac{1}{6}\left[\begin{array}{l} \frac{(r_3-r_1)^2 - (r_2-r_1)^2}{\|(r_3-r_1) \times (r_2-r_1)\|^2}[[(r_3-r_1) \times (r_2-r_1)] \times r_1] + \\ + \frac{(r_1-r_2)^2 - (r_3-r_2)^2}{\|(r_1-r_2) \times (r_3-r_2)\|^2}[[(r_1-r_2) \times (r_3-r_2)] \times r_2] + \\ + \frac{(r_2-r_3)^2 - (r_1-r_3)^2}{\|(r_2-r_3) \times (r_1-r_3)\|^2}[[(r_2-r_3) \times (r_1-r_3)] \times r_3] \end{array}\right].$$

(16)

Высота пирамиды определяется формулой:

$$h^2 = (R_{pr} + R_{sec})^2 - (r_1^2 + r_2^2 + r_3^2)/3 - p_c^2 + 2p_c \cdot \frac{(r_1 + r_2 + r_3)}{3},$$

(17)

где $R_{sec}$ - радиус сферы вторичной обкатки.

Смещение одного атома может приводить к смещению всех трех первичных опорных сфер. Производная центра шара-зонда определяется суперпозицией смещений, каждая из компонент которой соответствует сдвигу только одной из опорных сфер.

$$\frac{\partial r}{\partial R_{i,x^k}} = \frac{(r-r_1) \cdot \frac{\partial r_1}{\partial R_{i,x^k}}}{(r-r_1) \cdot [(r-r_2) \times (r-r_3)]}[(r-r_2) \times (r-r_3)] +$$

$$+ \frac{(r-r_2) \cdot \frac{\partial r_2}{\partial R_{i,x^k}}}{(r-r_2) \cdot [(r-r_1) \times (r-r_3)]}[(r-r_1) \times (r-r_3)] +$$

$$+ \frac{(r-r_3) \cdot \frac{\partial r_3}{\partial R_{i,x^k}}}{(r-r_3) \cdot [(r-r_2) \times (r-r_1)]}[(r-r_2) \times (r-r_1)].$$

(18)

Градиенты центров опорных первичных сфер, входящих в эту формулу, определяются ранее найденной формулой (13).

Отсюда для параметров поверхностного элемента:

$$\frac{\partial \boldsymbol{r}_s}{\partial R_{i,x^k}} = \frac{\partial \boldsymbol{r}}{\partial R_{i,x^k}},$$

$$\frac{\partial \boldsymbol{n}_s}{\partial R_{i,x^k}} = 0,$$

$$\frac{\partial S_s}{\partial R_{i,x^k}} = 0.$$

( 19 )

### 3.4 Точки на поверхности вторичных сфер устойчивых положений (Рис.7)

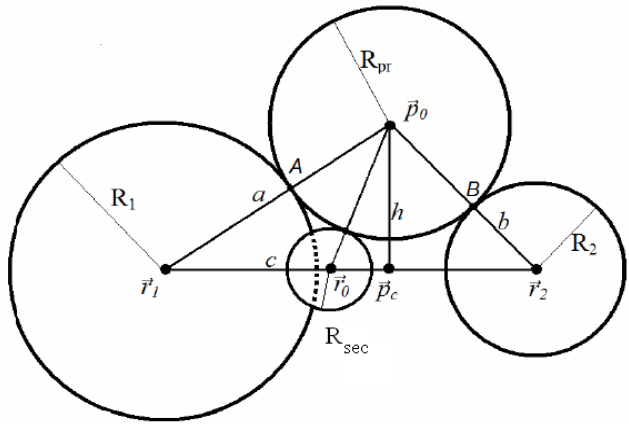

Рис. 7  Вторичные сферы устойчивых положений. См. [10].

Производные от параметров поверхностных элементов (координат, нормалей, площадей) для точек находящихся на поверхности сфер устойчивых положений полностью определяется градиентом центров этих сфер. Поэтому вначале вычислим градиенты центров этих сфер. Рассмотри треугольник, образованный центрами двух опорных атомов и любой из сфер первичной обкатки:

$$a = R_1 + R_{pr}; \quad b = R_2 + R_{pr}; \quad c = |\boldsymbol{R}_m - \boldsymbol{R}_n|,$$

( 20 )

$$a, b - const \quad \frac{\partial a}{\partial R_{i,x^k}} = \frac{\partial b}{\partial R_{i,x^k}} = 0,$$

( 21 )

$$\frac{\partial c}{\partial R_{i,x^k}} = \frac{(\delta_{in} - \delta_{im}) \cdot (R_{n,x^k} - R_{m,x^k})}{c},$$

( 22 )

*a, b, c* – стороны треугольника, $R_{pr}$ - радиус сферы первичной обкатки; $\boldsymbol{r}_1 = \boldsymbol{R}_m, \boldsymbol{r}_2 = \boldsymbol{R}_n$ - радиус-векторы опорных атомов.

Высота треугольника и ее градиент:

$$h = \frac{1}{2c}\sqrt{4a^2c^2 - (a^2 + c^2 - b^2)^2},$$

( 23 )

$$\frac{\partial h}{\partial R_{i,x^k}} = \left(\frac{b^2 + a^2 - c^2}{2h} - h\right) \cdot \frac{\frac{\partial c}{\partial R_{i,x^k}}}{c}.$$

( 24 )

Расстояние от центра сферы устойчивого положения до основания высоты треугольника и ее градиент:

$$d = \sqrt{(R_{pr} + R_{\sec})^2 - h^2},$$

( 25 )

$$\frac{\partial d}{\partial R_{i,x^k}} = -\frac{h \cdot \frac{\partial h}{\partial R_{i,x^k}}}{d},$$

( 26 )

где $R_{sec}$ - радиус сферы вторичной обкатки.

Единичный орт вдоль оси тора обкатки от первого атома ко второму и его градиент:

$$z = \frac{(R_n - R_m)}{c},$$

( 27 )

$$\frac{\partial z}{\partial R_{i,x^k}} = \frac{(\delta_{in} - \delta_{im})e^k}{c} - \frac{(R_n - R_m)}{c^2}\frac{\partial c}{\partial R_{i,x^k}}, \quad \left(\frac{\partial z}{\partial R_{i,x^k}} \cdot z\right) = 0.$$

( 28 )

Радиус вектор точки основания высоты треугольника $p_c$ и его градиент:

$$p_c = \tfrac{1}{2}(R_m + R_n) + \frac{(R_n - R_m)(a^2 - b^2)}{2c^2},$$

( 29 )

$$\frac{\partial p_c}{\partial R_{i,x^k}} = \tfrac{1}{2}(\delta_{in} + \delta_{im})\vec{e}^k + \frac{(\delta_{in} - \delta_{im})}{2}\frac{(a^2 - b^2)}{c^2}\vec{e}^k - (R_n - R_m)(a^2 - b^2)\frac{((\delta_{in} - \delta_{im}) \cdot (R_{n,x^k} - R_{m,x^k}))}{c^4}.$$

( 30 )

$r_0$ - радиус-вектора центров двух сфер устойчивого положения и их градиенты:

$$r_0 = p_c \pm dz,$$

$$\frac{\partial r_0}{\partial R_{i,x^k}}_0 = \frac{\partial p_c}{\partial R_{i,x^k}} \pm \left(\frac{\partial d}{\partial R_{i,x^k}} \cdot z + d \cdot \frac{\partial z}{\partial R_{i,x^k}}\right).$$

( 31 )

Отсюда для параметров поверхностного элемента:

$$\frac{\partial \boldsymbol{r}_s}{\partial R_{i,x^k}} = \frac{\partial \boldsymbol{r}_0}{\partial R_{i,x^k}},$$

$$\frac{\partial \boldsymbol{n}_s}{\partial R_{i,x^k}} = 0,$$

$$\frac{\partial S_s}{\partial R_{i,x^k}} = 0.$$

( 32 )

### 3.5 Первичная обкатка двух атомов. Торы первичной обкатки.(Рис.8)

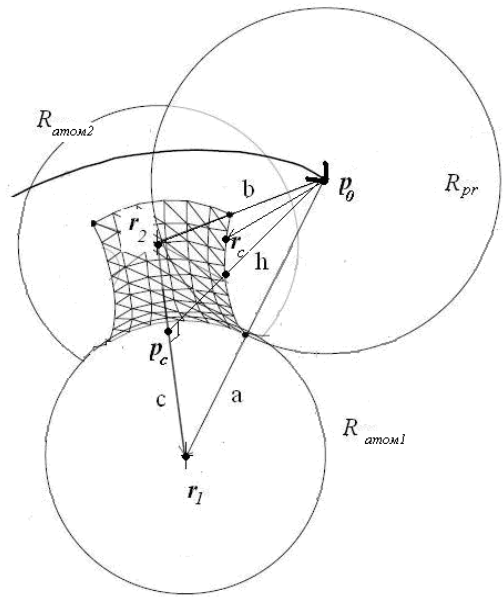

Рис. 8 Торы первичной обкатки при опоре на два атома.

Частично необходимые формулы уже выведены в предыдущем пункте. Пусть $x, y, z$ - локальный базис тора (центр координат лежит в основании высоты). Ось z определена как в предыдущем пункте

$(\boldsymbol{x}\cdot\boldsymbol{z})=0,\ (\boldsymbol{y}\cdot\boldsymbol{z})=0,\ (\boldsymbol{x}\cdot\boldsymbol{y})=0,\ (\boldsymbol{x}\cdot\boldsymbol{x})=1,\ (\boldsymbol{y}\cdot\boldsymbol{y})=1,\ (\boldsymbol{z}\cdot\boldsymbol{z})=1$.

( 33 )

Нормаль поверхностного элемента разлагается по локальному базису:

$\boldsymbol{n}_s = \alpha_s\sqrt{1-\gamma_s^2}\,\boldsymbol{x} + \beta_s\sqrt{1-\gamma_s^2}\,\boldsymbol{y} + \gamma_s \boldsymbol{z},$

( 34 )

$$\alpha_s{}^2 + \beta_s{}^2 = 1,$$

( 35 )

$$\frac{\partial \gamma_s}{\partial R_{i,x^k}} = 0, \quad \frac{\partial \alpha_s}{\partial R_{i,x^k}} = 0, \quad \frac{\partial \beta_s}{\partial R_{i,x^k}} = 0 \quad .$$

( 36 )

Задания $\alpha_s$, $\beta_s$, $\gamma_s$ определяют положение центра поверхностного элемента на торе.

$e_s$ -единичный вектор из основания высоты к центру сферы обкатки, определяемой текущими $\alpha_s, \beta_s$

$$e_s = \alpha_s x + \beta_s y.$$

( 37 )

Его градиент:

$$\frac{\partial e_s}{\partial R_{i,x^k}} = -\left(e_s \cdot \frac{\partial z}{\partial R_{i,x^k}}\right) z \quad .$$

( 38 )

Отсюда градиент нормали поверхностного элемента для торов первичной обкатки:

$$\frac{\partial n_s}{\partial R_{i,x^k}} = \sqrt{1-\gamma_s^2}\, \frac{\partial e_s}{\partial R_{i,x^k}} + \gamma_s \cdot \frac{\partial z}{\partial R_{i,x^k}} \,.$$

( 39 )

Для градиента радиус-вектора поверхностного элемента:

$$\frac{\partial r_s}{\partial R_{i,x^k}} = \frac{\partial p_c}{\partial R_{i,x^k}} + \frac{\partial h}{\partial R_{i,x^k}} \cdot e_s + \left(h - R_{rp}\sqrt{1-\gamma_s^2}\right) \cdot \frac{\partial e_s}{\partial R_{i,x^k}} - R_{rp}\gamma_s \cdot \frac{\partial z}{\partial R_{i,x^k}}.$$

( 40 )

Для градиента площадей поверхностных элементов торов первичной обкатки:

$$\frac{\partial S_s}{\partial R_{i,x^k}} = \frac{\frac{\partial h}{\partial R_{i,x^k}}}{(h - R_{rp}\sqrt{1-\gamma_s^2})} \cdot S_s .$$

( 41 )

### 3.6 Вторичная обкатка двух сфер первичной обкатки. Торы вторичной обкатки.(Рис.9)

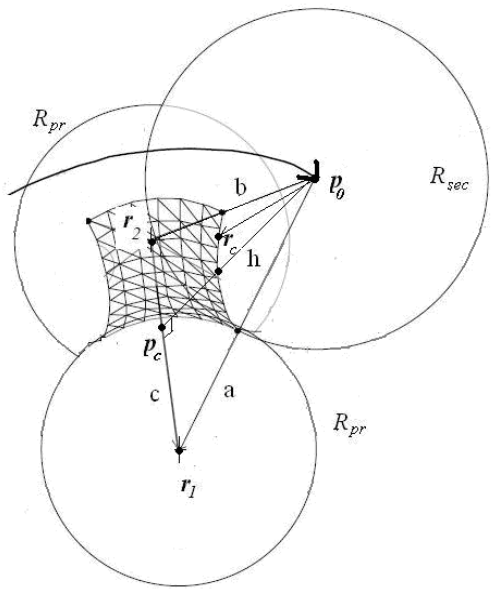

Рис. 9 Торы вторичной обкатки при опоре на две первичные сферы.

Этот случай похож на предыдущий. Он даже проще – опорные сферы имеют одинаковый радиуса и треугольник, образованный центрами опорных сфер и сферой вторичной обкатки является равносторонним. Однако есть и отличие – все нормали направлены в направлении, *противоположным* центру текущей сферы обкатки.

Стороны тругольника, образованного центрами двух сфер первичной обкатки и сферой вторичной обкатки и их градиенты:

$$a = b = R_{pr} + R_{sec}; \quad c = |\boldsymbol{r}_2 - \boldsymbol{r}_1|, \qquad (42)$$

$$\frac{\partial c}{\partial R_{i,x^k}} = \frac{(\frac{\partial \boldsymbol{r}_2}{\partial R_{i,x^k}} - \frac{\partial \boldsymbol{r}_1}{\partial R_{i,x^k}}) \cdot (\boldsymbol{r}_2 - \boldsymbol{r}_1)}{c}. \qquad (43)$$

Высота и ее градиент:

$$h = \frac{1}{2}\sqrt{4a^2 - c^2}, \qquad (44)$$

$$\frac{\partial h}{\partial R_{i,x^k}} = -\frac{c}{4h} \cdot \frac{\partial c}{\partial R_{i,x^k}}. \qquad (45)$$

Радиус вектор точки основания высоты треугольника $\boldsymbol{p}_c$ и ее производная:

$$\boldsymbol{p}_c = \tfrac{1}{2}(\boldsymbol{r}_1 + \boldsymbol{r}_2),$$
$$\frac{\partial \boldsymbol{p}_c}{\partial R_{i,x^k}} = \tfrac{1}{2}\left( \frac{\partial \boldsymbol{r}_1}{\partial R_{i,x^k}} + \frac{\partial \boldsymbol{r}_2}{\partial R_{i,x^k}} \right). \qquad (46)$$

$\boldsymbol{x}, \boldsymbol{y}, \boldsymbol{z}$ - локальный базис

$$\boldsymbol{z} = \frac{(\boldsymbol{r}_2 - \boldsymbol{r}_1)}{c}, \quad (\boldsymbol{x} \cdot \boldsymbol{z}) = 0, \quad (\boldsymbol{y} \cdot \boldsymbol{z}) = 0, \quad (\boldsymbol{x} \cdot \boldsymbol{y}) = 0. \qquad (47)$$

Производные локального базиса

$$\frac{\partial z}{\partial R_{i,x^k}} = \frac{\left(\frac{\partial r_2}{\partial R_{i,x^k}} - \frac{\partial r_1}{\partial R_{i,x^k}}\right)}{c} - \frac{(r_2 - r_1)}{c^2}\frac{\partial c}{\partial R_{i,x^k}}.$$

( 48 )

Проекция нормали на локальный базис для торов вторичной обкатки (нормаль направлена внутрь тора):

$$n_s = -\alpha_s \cdot \sqrt{1-\gamma_s^2} \cdot x - \beta_s \cdot \sqrt{1-\gamma_s^2} \cdot y - \gamma_s \cdot z.$$

( 49 )

Производные радиус-вектора центра текущей сферы обкатки, единичного вектора из основания высоты к центру текущей сферы обкатки, векторов локального базиса $x, y, z$:

$$e_s = \alpha_s x + \beta_s y,$$

( 50 )

$$\frac{\partial e_s}{\partial R_{i,x^k}} = -\left(e_s \cdot \frac{\partial z}{\partial R_{i,x^k}}\right) z.$$

( 51 )

Отсюда градиент нормали поверхностного элемента для торов вторичной обкатки:

$$\frac{\partial n_s}{\partial R_{i,x^k}} = -\sqrt{1-\gamma_s^2}\frac{\partial e_s}{\partial R_{i,x^k}} - \gamma_s \cdot \frac{\partial z}{\partial R_{i,x^k}}.$$

( 52 )

Для градиента радиус-вектора поверхностного элемента:

$$\frac{\partial \boldsymbol{r}_s}{\partial R_{i,x^k}} = \frac{\partial \boldsymbol{p}_c}{\partial R_{i,x^k}} + \frac{\partial h}{\partial R_{i,x^k}} \cdot \boldsymbol{e}_s + \left(h - R_{sec}\sqrt{1-\gamma_s^2}\right) \cdot \frac{\partial \boldsymbol{e}_s}{\partial R_{i,x^k}} - R_{sec}\gamma_s \cdot \frac{\partial \boldsymbol{z}}{\partial R_{i,x^k}}.$$

( 53 )

Для градиента площадей поверхностных элементов торов вторичной обкатки:

$$\frac{\partial S_s}{\partial R_{i,x^k}} = \frac{\frac{\partial h}{\partial R_{i,x^k}}}{(h - R_{sec}\sqrt{1-\gamma_s^2})} \cdot S_s.$$

( 54 )

### 3.7 Рассмотрим граничный многоугольный поверхностный элемент, который состоит из треугольников, чьи вершины не принадлежат одному и тому же поверхностному фрагменту.

Нормали и координаты радиус-вектора многоугольного поверхностного элемента определяются типом фрагмента поверхности, на которой лежит его центр. По формулам, относящимся к этому фрагменту, вычисляются и градиенты радиус-вектора и нормалей этого поверхностного элемента.

С его площадью сложнее. Ее величина равна трети суммы площадей составляющих его треугольников. Соответственно, ее градиент

$$\frac{\partial S^M}{\partial R_{i,x^k}} = \frac{\sum_j \frac{\partial S_j^{tr}}{\partial R_{i,x^k}}}{3}.$$

( 55 )

Пусть треугольник образован вершинами $\boldsymbol{r}_1, \boldsymbol{r}_2, \boldsymbol{r}_3$

$$\boldsymbol{a}_j = \boldsymbol{r}_2 - \boldsymbol{r}_1$$
$$\boldsymbol{b}_j = \boldsymbol{r}_3 - \boldsymbol{r}_1.$$

( 56 )

Его площадь

$$s = [a_j \times b_j], \quad (57)$$

$$S_j^{tr} = \frac{1}{2}|s|. \quad (58)$$

Градиент его площади:

$$\frac{\partial S_j^{tr}}{\partial R_{i,x^k}} = \frac{1}{4} \frac{(\frac{\partial a_j}{\partial R_{i,x^k}} \cdot a_j)b_j^2 + (\frac{\partial b_j}{\partial R_{i,x^k}} \cdot b_j)a_j^2 - (a_j \cdot b_j)((\frac{\partial a_j}{\partial R_{i,x^k}} \cdot b_j) + (a_j \cdot \frac{\partial b_j}{\partial R_{i,x^k}}))}{S_j^{tr}}, \quad (59)$$

$$\frac{\partial a_j}{\partial R_{i,x^k}} = \frac{\partial r_2}{\partial R_{i,x^k}} - \frac{\partial r_1}{\partial R_{i,x^k}},$$

$$\frac{\partial b_j}{\partial R_{i,x^k}} = \frac{\partial r_3}{\partial R_{i,x^k}} - \frac{\partial r_1}{\partial R_{i,x^k}}. \quad (60)$$

Мы предположили выше, что градиенты нормалей граничных элементов могут рассчитываться на основе градиента нормали к поверхности, проходящей через центр этого элемента. Однако можно посчитать это величину и иначе. Многоугольный поверхностный элемент состоит из треугольных граней. Можно посчитать нормаль как среднюю от нормалей всех ее граней (взвешенную по площадям). На основе этого затем посчитать и производную нормали, что мы и сделаем ниже. Практические расчеты показали, что эти производные приводят к меньшим численным ошибкам в градиенте.

Однако для самой энергии меньшие ошибки получаются для нормалей, взятых в центре поверхностного элемента.

Это все связанно с тем, что многоугольные элементы состоят из треугольников, относящихся к разным поверхностным фрагментам, Их площади сильно меняются и вершины сильно сдвигаются *после* искажения поверхностной сетки сдвигом атома, по которому берется градиент. Поэтому и средняя нормаль сильно отличается от нормали в центре *после* искажения поверхностной сетки сдвигом атома.

Опишем пересчет градиентов нормалей граничных поверхностных элементов, описанный выше.

Нормаль элемента ищем как векторную среднюю от нормалей к треугольным элементам с вершиной в центральной точке многоугольного элемента. Веса пропорциональны площадям треугольников.

$$a_{gi} = r_{gi} - r_g,$$

( 61 )

где

$r_g$ – радиус-вектор центра текущего поверхностного элемента

$r_{gi}$ – радиус-вектора центров соседних поверхностных элементов.

$i=1,...,N_{eg}$, $N_{eg}$ – число соседних поверхностных элементов.

$n_g$ - прежняя нормаль в центре поверхностного элемента

$$a_{gi\_next} = \begin{cases} a_{g(i+1)} & \text{for } 1 \leq i < N_{eg} \\ a_{g1} & \text{for } i = N_{eg} \end{cases}.$$

( 62 )

Новая нормаль:

$$n_{g(new)} = \frac{\sum_{i=1}^{N_{eg}} \left( [a_{gi} \times a_{gi\_next}] sign(n_g [a_{gi} \times a_{gi\_next}]) \right)}{\left| \sum_{i=1}^{N_{eg}} \left( [a_{gi} \times a_{gi\_next}] sign(n_g [a_{gi} \times a_{gi\_next}]) \right) \right|}.$$

( 63 )

Градиент нормали считаем через градиенты вершин треугольников по формуле, аналогичной формуле для градиента площади граничного поверхностного элемента.

$$L_{gi} = sign\left( n_g \cdot [a_{gi} \times a_{gi\_next}] \right),$$  ( 64 )

$$s_g = \sum_i L_{gi}\left[a_{gi} \times a_{gi\_next}\right], \qquad (65)$$

$$\frac{\partial a_{gi}}{\partial R_{m,x^k}} = \frac{\partial r_{gi}}{\partial R_{m,x^k}} - \frac{\partial r_g}{\partial R_{m,x^k}},$$

(66)

$$\frac{\partial a_{gi\_next}}{\partial R_{m,x^k}} = \frac{\partial r_{gi\_next}}{\partial R_{m,x^k}} - \frac{\partial r_g}{\partial R_{m,x^k}},$$

(67)

$$S_{gi} = \left(\left[\frac{\partial a_{gi}}{\partial R_{m,x^k}} \times a_{gi\_next}\right] + \left[a_{gi} \times \frac{\partial a_{gi\_next}}{\partial R_{m,x^k}}\right]\right)L_{gi}, \qquad (68)$$

$$\frac{\partial n_{g(new)}}{\partial R_{m,x^k}} = \frac{\sum_{i=1}^{N_{eg}}\left(S_{gi} - n_{g(new)}\left(n_{g(new)} \cdot S_{gi}\right)\right)}{|\vec{S}_g|}. \qquad (69)$$

### 4. Поверхностные элементы на SAS и их аналитические градиенты.

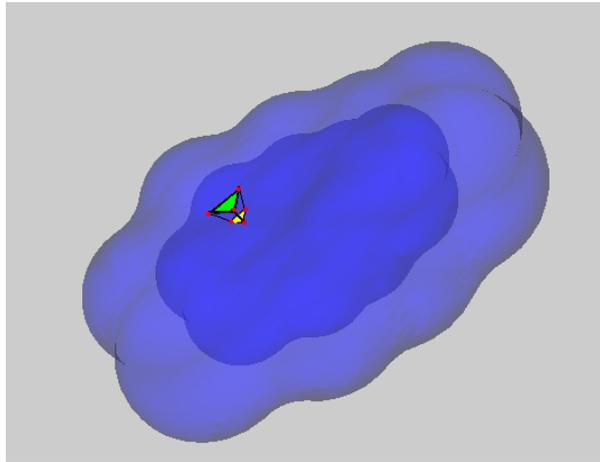

Рис. 10. Преобразование поверхности типа SES в поверхность типа SAS.

Пусть имеется поверхность SES после первичной обкатки.

Пусть имеется треугольник на первичной поверхности SES ($r_1$,$r_2$,$r_3$)

Образ этого треугольника на SAS ($r_{1n}$,$r_{2n}$,$r_{3n}$)(Рис.10)

$$r_{1n} = r_1 + n_1 R_{pr},$$
( 70 )

$$r_{2n} = r_2 + n_2 R_{pr},$$
( 71 )

$$r_{3n} = r_3 + n_3 R_{pr},\qquad(72)$$

где $R_{pr}$ – радиус первичной обкатки, $n_1$, $n_2$, $n_3$ - нормали в соответствущий точках.

1) поверхностный элемент на первичной сфере, опирающейся на три атома, отображается в точку. Его градиент площади нулевой
2) поверхностный элемент на торе, опирающимся на два атома, отображается в линию. Его градиент площади нулевой
3) поверхностный элемент на атоме отображается в сферический тругольник. Градиент площади по смещению атомов равен нулю
4) граничный поверхностный элемент между тором и первичной сферой, опирающейся на три атома. Имеет нулевую площадь и, следовательно, его градиент площади нулевой
5) граничный треугольный поверхностный элемент, опирающийся на первичный тор и атом, первичную сферу и атом или первичные тор, сферу и атом – площадь считается по формуле (58) для трех точек ($r_{1n}$, $r_{2n}$, $r_{3n}$). Градиент его площади считаем по формуле:

$$\frac{\partial s_j}{\partial R_{m,x^k}} = \frac{(b \cdot b)\left(\frac{\partial a}{\partial R_{m,x^k}} \cdot a\right) + (a \cdot a)\left(\frac{\partial b}{\partial R_{m,x^k}} \cdot b\right) - (a \cdot b)\left(\left(\frac{\partial a}{\partial R_{m,x^k}} \cdot b\right) + \left(\frac{\partial b}{\partial R_{m,x^k}} \cdot a\right)\right)}{S_4},\qquad(73)$$

где стороны треугольного поверхностного элемента и их градиенты:

$$a = r_{2n} - r_{1n},\qquad(74)$$

$$\boldsymbol{b} = \boldsymbol{r}_{3n} - \boldsymbol{r}_{1n},$$
( 75 )

$$\frac{\partial \boldsymbol{a}}{\partial R_{m,x^k}} = \frac{\partial \boldsymbol{r}_{2n}}{\partial R_{m,x^k}} - \frac{\partial \boldsymbol{r}_{1n}}{\partial R_{m,x^k}},$$
( 76 )

$$\frac{\partial \boldsymbol{b}}{\partial R_{m,x^k}} = \frac{\partial \boldsymbol{r}_{3n}}{\partial R_{m,x^k}} - \frac{\partial \boldsymbol{r}_{1n}}{\partial R_{m,x^k}},$$
( 77 )

$$S_4 = 2|\boldsymbol{a} \times \boldsymbol{b}|,$$
( 78 )

градиенты $\boldsymbol{r}_{1n}, \boldsymbol{r}_{2n}, \boldsymbol{r}_{3n}$:

$$\frac{\partial \boldsymbol{r}_{1n}}{\partial R_{m,x^k}} = \frac{\partial \boldsymbol{r}_{1n}}{\partial R_{m,x^k}} + \frac{\partial \boldsymbol{n}_1}{\partial R_{m,x^k}} R_{pr},$$
( 79 )

$$\frac{\partial \boldsymbol{r}_{2n}}{\partial R_{m,x^k}} = \frac{\partial \boldsymbol{r}_{2n}}{\partial R_{m,x^k}} + \frac{\partial \boldsymbol{n}_2}{\partial R_{m,x^k}} R_{pr},$$
( 80 )

$$\frac{\partial \boldsymbol{r}_{3n}}{\partial R_{m,x^k}} = \frac{\partial \boldsymbol{r}_{3n}}{\partial R_{m,x^k}} + \frac{\partial \boldsymbol{n}_3}{\partial R_{m,x^k}} R_{pr}.$$
( 81 )

Индекс $j=1,..,N$ пробегает по всем треугольным поверхностным элементам с ненулевым градиентом на SAS.

## 5. Компьютерная проверка расчетов.

Была проведена компьютерная проверка найденных выше формул с помощью программы DISOLV [4-5,8-10]:

1) Строились две поверхности обкатки – исходная поверхность молекулы и поверхность, полученная при малом сдвиге одного из атомов. Сдвиг узлов сетки рассчитывался с помощью найденных выше градиентов координат поверхностных элементов путем домножения их на указанный сдвиг атома. Затем эти узлы проецировались на поверхность, полученную после сдвига атома. Таким образом, получались две новые сетки – сдвинутая и ее проекция на новую поверхность. При правильных градиентах разница в положении узлов этих двух сеток должна была быть мала, много меньше разницы между ними и исходной сеткой на исходной молекулярной поверхности. Кроме того, уменьшение этой разницы между двумя новыми сетками должно происходить нелинейно (и более быстро, чем линейно) при уменьшении сдвига атома, поскольку определяется вторыми производными. Все это действительно наблюдалось при компьютерных расчетах. Это подтверждает верность формул для производных координат поверхностных элементов.

2) Искались численные градиенты площадей поверхностных элементов через разницу площадей поверхностных элементов на старой и полученной новой сетке. Эта разница сравнивалась с градиентом площади поверхностного элемента, домноженного на сдвиг. Хорошее совпадение результатов подтверждает верность формул для производных площадей поверхностных элементов.

3) Делалось все то же самое, что и в пункте 1). Но кроме этого новая спроецированная сетка сдвигалась на величину нормали к новой поверхности. Новая сетка, полученная до проецирования тоже сдвигалась на длину пересчитанной нормали (старая нормаль плюс ее градиент, умноженный на сдвиг атома). Эти две новые сдвинутые сетки сравнивались между собой по тем же принципам, что изложены в пункте 1).

## 6. Выводы.

В работе проведены расчеты аналитических градиентов параметров поверхностных элементов молекулярной поверхности, полученной методами первичной и вторичной обкатки в соответствии с алгоритмами, изложенными в [3-10].

Приведем проблемные случаи, которые могут возникать при расчете градиентов и решении задачи оптимизации геометрии молекул с их использованием:

- Вырождение - тройная точка опирается не на три, а на большее число атомов (например, ароматические кольца)
- Происходит резкая перестройка поверхности из-за изменения радиуса вторичной обкатки или критического расстояния самой программой при автоматической его настройке
- Происходит резкая перестройка поверхности из-за появления или исчезновении тора обкатки (первичного или вторичного) при сдвиге тройных точек или изменения узкого перешейка тора на расстояниях близких к критическому
- Происходит резкая перестройка поверхности из-за появления или исчезновении тройных точек первичных или вторичных.

Как результат могут быть скачки градиентов и «колебания» оптимизатора вблизи вышеописанных точек

Компьютерный расчет дает хорошее совпадение с ожидаемыми результатами и подтверждает верность формул для градиентов нормалей поверхностных элементов.

## Благодарности.

Данная работа является более детальным изложением и дальнейшим развитием статьи [10]. Авторы выражают глубокую признательность всем авторам этой публикации за проделанную работу, которая послужила основой и для данной статьи.